\documentclass[
  journal=pasa,
  manuscript=research-paper, 
  year=202X,
  volume=YY,
]{cup-journal}

\usepackage{amsmath}
\usepackage{amssymb,microtype,siunitx,booktabs}
\usepackage{hyperref}
\newcommand{\Msun}[0]{$M_{\odot}$}
\newcommand{\Lsun}[0]{$L_{\odot}$}
\newcommand{\alphamlt}[0]{$\alpha_{\text{MLT}}$}

\title{Constraining intrinsic S-type AGB masses and third dredge-up with pulsation}

\author{Y. L. Mori}
\affiliation{School of Physics and Astronomy, Monash University, Clayton, Vic 3800, Australia}
\email[Yoshiya L. Mori]{yoshiya.mori@monash.edu}

\author{A. I. Karakas}
\affiliation{School of Physics and Astronomy, Monash University, Clayton, Vic 3800, Australia}

\author{S. W. Campbell}
\affiliation{School of Physics and Astronomy, Monash University, Clayton, Vic 3800, Australia}

\received {13 Aug 2025}
\revised  {16 Oct 2025}
\accepted {21 Oct 2025}
\published{dd Mmm YYYY}


\begin{document}

\begin{abstract}
The lowest mass at which the third dredge-up (TDU) occurs for thermally-pulsing asymptotic giant branch (TP-AGB) stars remains a key uncertainty in detailed stellar models. S-type AGB stars are an important constraint on this uncertainty as they have C/O ratios between 0.5 and 1, meaning they have only experienced up to a few episodes of TDU. AGB stars are also long-period variable stars, pulsating in low order radial pulsation modes. 

In this paper we estimate the initial masses of a large sample of intrinsic S-type AGB stars, by analysing their visual light curves, estimating their luminosities with Gaia DR3 parallax distances and finally comparing to a grid of detailed stellar models combined with linear pulsation models. 

We find that the initial mass distribution of intrinsic S-type stars peaks at 1.3 to 1.4 \Msun, depending on model assumptions. There also appear to be stars with initial masses down to 1 solar mass, which is in conflict with current detailed stellar models. Additionally, we find that though the mass estimates for semiregular variable stars pulsating in higher order radial modes are precise, the Mira variables pulsating in the fundamental mode present challenges observationally from uncertain parallax distances, and theoretically from the onset of increased mass-loss and the necessity of non-linear pulsation models.\end{abstract}

\section{INTRODUCTION }
\label{sec:int}

Low- to intermediate-mass ($0.8-8$\Msun) stars on the asymptotic giant branch (AGB) will undergo repeated helium shell flash events, also known as thermal pulses (TPs) \citep{karakas_dawes_2014}. These TPs drive convective mixing between the envelope and the intershell, in which the products of partial helium burning such as carbon ($^{12}$C) are brought to the stellar surface in a process known as the third dredge-up (TDU). These TDU events can gradually alter the stellar surface composition of the AGB star from being oxygen rich (C/O $<$ 0.5) to carbon rich (C/O $>$ 1), which corresponds to a change in spectral type of M-type to C(N)-type. 

In detailed models of AGB evolution, the lowest initial mass at which TDU can occur remains uncertain. This uncertainty is mainly due to the numerical treatment of the uncertain physics of convective boundary mixing \cite[e.g.][]{frost_numerical_1996}. A key observational constraint on the minimum mass for TDU is the carbon star luminosity function of the Magellanic Clouds, which have been used to demonstrate the qualitative trend of higher TDU efficiency at lower metallicities \citep{groenewegen_synthetic_1993}. However, these early studies also suggest that TDU should occur at masses as low as 1.2-1.4 \Msun\ \citep{marigo_third_1999}. This is a challenge even for current theoretical models, as they require a significant addition of convective overshoot to induce TDU even for models of 1.5 \Msun\ \citep{herwig_overshoot_2000,cristallo_agb_2009,kamath_evolution_2012,karakas_stellar_2016}. 

Stars of spectral type S are identified by their strong ZrO absorption features, and have C/O between 0.5 to 0.99 \citep{keenan_classification_1954,piccirillo_interpretation_1980}. S-type AGB stars are of particular interest for constraining TDU as they represent an intermediate stage of surface carbon enrichment before potentially becoming carbon-rich, following the progression from spectral type M $\rightarrow$ S $\rightarrow$ SC (C/O $\sim1$) $\rightarrow$ C(N). They are further classified into intrinsic and extrinsic S-type stars, based on whether they are self-enriched in carbon (intrinsic) or enriched by mass transfer from a former AGB companion that is now a white dwarf (extrinsic). There are a number of methods to differentiate between the two, such as colour-colour diagrams \citep{yang_infrared_2006} and radial velocity measurements to detect a companion \citep{van_eck_henize_2000b}, but the strongest evidence of recent TDU in an S-type star is the detection of the s-process element technetium (Tc). Its isotope, $^{99}$Tc, has a half-life of $\sim210,000$ years, which is relatively short compared to the main-sequence lifetime -- so its existence is good evidence that there is ongoing s-process element nucleosynthesis. 

AGB stars are also pulsating long-period variable (LPV) stars, which have variability periods on the order of tens to thousands of days. LPVs include the Mira variables which have visual amplitudes $>2.5$ mag, semiregular variables (SRV) which show multiple variability periods and the OGLE small amplitude red giants (OSARGs). These stars have been found to form distinct sequences in the period-luminosity (PL) diagram, most clearly established in studies of LPVs in the Magellanic Clouds \citep{wood_macho_1999,soszynski_optical_2005}. The sequences A, A$'$, B, C$'$ and C have been associated with low-order radial mode pulsation, while the origin of the long secondary periods (LSPs) in sequence D is still debated \citep{nicholls_long_2009,takayama_lsp_2020,soszynski_binarity_2021,goldberg_betelbuddy_2024}. The use of PL relations of Mira variables to determine distances has been well-studied in existing studies \citep{whitelock_agb_2008}, as well as for the semiregular variables \citep{trabucchi_semi-regular_2021,hey_dists_2023}. 

The modelling of AGB pulsation has had sustained effort motivated by the aforementioned observations. \cite{fox_theoretical_1982} computed linear, non-adiabatic pulsation model grids of red giant stars, which form the basis for the models in \cite{wood_strange_2014} and later \cite{trabucchi_modelling_2019} (hereafter T19). These models conduct linear stability analysis on spherically symmetric, static envelopes of AGB stars, and calculate the pulsation periods for given stellar parameters. The T19 models specifically target typical stellar parameters for TP-AGB stars, and also account for the changes in opacity due to chemical enrichment from TDU. Similar studies include that of \cite{xiong_non-adiabatic_2007} and \cite{xiong_turbpulsation_2018} who used a non-local, time-dependent theory of convection in their models as opposed to the commonly used mixing length theory. Aside from 1D models of AGB pulsation, there has also been recent work in 3D models which consider the complex multidimensional interactions between large-scale convection and pulsation \citep{freytag_3dhydroagb_2017,ahmad_3dpulsation_2023,ahmad_multimode_2025}.

The study of pulsation and its relationship with mass-loss on the AGB is also a key component in our understanding of their evolution, as mass-loss essentially governs the final stages and termination of the AGB phase. Studies such as \cite{uttenthaler_pmasslossmiras_2013} and \cite{uttenthaler_interplay_2019} have analysed the behaviour of AGB variables in photometric colour and period, finding that the Tc-rich and Tc-poor Mira variables form separate sequences in the period vs. K - [22] colour diagram.

There have been several studies that estimate masses for AGB stars using their pulsation, with some also placing constraints on TDU using their results. \cite{takeuti_method_2013} compared measurements of AGB stars with the models of \cite{xiong_non-adiabatic_2007} to estimate their current masses, including stars pulsating in multiple modes. \cite{kamath_pulsation_2010} investigated a collection of AGB variable stars in two Magellanic Cloud Clusters, finding masses in good agreement with isochrones. These results were used to further constrain AGB models, finding that significant convective overshoot (1-3 pressure scale heights) is required to reproduce the observed O-rich to C-rich transition \citep{kamath_evolution_2012}. \cite{marigo_fresh_2022} examined AGB stars in open clusters, using their well-known distances and measuring their pulsation periods to constrain TDU. Their sample was limited to three MS and S-type stars, deriving masses of 1.6, 2.1 and 2.8 \Msun\ for these objects. Studying large samples of S-type stars in clusters can be challenging, given their transitive nature in an already short-lived stage of stellar evolution and the need to ascertain whether or not they are intrinsic or extrinsic. 

Recent studies have specifically investigated the stellar parameters and chemistry of intrinsic S-type AGB stars. \cite{abia_characterisation_2022} use Gaia DR3 parallax distances to investigate the luminosity functions of C(N)-type and related stars. This study found that the intrinsic S stars have average M$_{\text{bol}} = -4.42 \pm 0.68$ mag, finding a lower luminosity limit of M$_{\text{bol}} \sim -4.0$ mag, which corresponds to initial masses of 1.3 \Msun\ at solar metallicity. Another key collection of papers is \cite{shetye_s_2018,shetye_observational_2019,shetye_s_2021}; especially \citealp{shetye_s_2021}, hereafter S21, which use high resolution spectra and Gaia parallax of S-type AGB stars to derive important stellar parameters such as luminosity, effective temperature, surface gravity, C/O and metallicity. Interestingly, some of the stars in this study were estimated to have initial masses below 1.5 \Msun\ and even as low as 1 \Msun, which presents a further challenge to the initial mass at which TDU occurs in stellar evolution calculations. Another interesting result of this study was that a disproportionate number of initial masses in the paper appear to be quite high ($\sim3$ \Msun) given a random sample of TP-AGB stars. These masses were estimated from the Hertzsprung-Russell diagram, which can be somewhat challenging given the long-term (thermal pulse cycle) and shorter term (pulsation) variability of these stars. Of course, these limitations were perhaps beyond the focus of the study which was on high quality measurements for a smaller sample of stars.

In this paper, we aim to derive accurate masses for a large collection of intrinsic S-type stars using their pulsation characteristics and luminosities. The paper structure is as follows: in Section \ref{s:literature_sample} we describe the sample of intrinsic S-type stars drawn from the literature. Next, in Section \ref{s:light_curves} we describe our method in which we analyse the light curves from the ASAS-SN survey to measure pulsation periods for the sample, and then derive their luminosities in Section \ref{s:luminosities}. We then compare these derived parameters with current models of stellar evolution and pulsation on the TP-AGB described in Section \ref{s:pulsation_models_tpagb}, to derive pulsation masses for the sample in Section \ref{s:results}. Finally, we discuss these results in Section \ref{s:discussion} and conclude in Section \ref{s:conclusion}.

\section{SAMPLE SELECTION}
\label{s:literature_sample}

In order to collect a clean sample of intrinsic S-type stars from existing studies, we must clearly indicate the method with which a star has been classified as intrinsic. Table \ref{table:sample_overview} summarises the papers from which we collect our sample of intrinsic S-type stars, as well as counts after applying several quality cuts described in this section. We collect S stars that have both Tc detections in published results, which we denote as `Tc-yes' stars, and S stars that have been classified as intrinsic via other means as `Tc-maybe' stars.

We will first briefly summarise the papers which have both Tc-yes and Tc-maybe stars, with a few different approaches to classifying stars as intrinsic. \cite{van_eck_henize_2000a} and \cite{van_eck_henize_2000b} analysed a sample of S stars from the Henize catalogue of S stars \citep{henize_sstars_1960}, and classify them as intrinsic or extrinsic. Their intrinsic classifications are based on either the Tc detections in \cite{van_eck_jorissen_tc_1999}, or with a multivariate classification scheme which uses an unsupervised clustering algorithm taking into account parameters of radial velocity measurements, UBV and IRAS K \& [12] photometry, and low resolution spectroscopy (see paper for full details). \cite{yang_infrared_2006} classify intrinsic and extrinsic S stars using colour-colour diagrams with 2MASS, IRAS and MSX photometry, such as those in \cite{jorissen_tc_1993}, \cite{chen_irsstars_1998} and \cite{wang_chen_nir_2002}. Both of the above two classifications are used in \cite{guandalini_infrared_2008}, and are carried through to the \cite{chen_infrared_2019} sample of known intrinsic S stars. We finally include the intrinsic S stars that were newly classified in \cite{chen_infrared_2019}, which also makes classifications based on colour-colour diagrams with 2MASS, WISE and IRAS photometry. From this sample, Hen 4-255 was corrected to Hen 4-225, and DR CMa to DK CMa using their GCSS numbers.

We also cross-match with the samples of \cite{uttenthaler_interplay_2019} and \cite{shetye_s_2018,shetye_observational_2019,shetye_s_2021} which limit their sample to AGB stars that have confirmed Tc measurements in prior work. From the Tc-rich stars in \cite{uttenthaler_interplay_2019}, we also include those with spectral types MS and M. \cite{uttenthaler_interplay_2019} also has a comprehensive list of references for the stars with Tc measurements. The sample from \cite{shetye_s_2018,shetye_observational_2019,shetye_s_2021} all have reliable Tc measurements from high-resolution spectra, with additional estimates of stellar parameters such as T$_{\text{eff}}$, surface gravity log g, metallicity ([Fe/H]), C/O and [s/Fe]\footnote{$[A/B] = \log_{10}(A/B)_{\text{surf}} - \log_{10}(A/B)_{\odot}$, where $(A/B)_{\text{surf}}$ is the number ratio of elements A and B at the stellar surface and $(A/B)_{\odot}$ is the solar number ratio.}.

\begin{table*}
    \centering
    \caption{Counts of intrinsic S-type stars from the published studies, and the sample after quality cuts. Additional cuts refer to the removal of stars with new Tc-poor classifications, and extrinsic categorisation from later sections. The final sample is comprised of stars that have been classified as intrinsic via Tc detections, other intrinsic classifications, have acceptable ASAS-SN light curves, pass the parallax error cut and finally have one or more reliably measured variability periods.}
    \label{table:sample_overview}
        \begin{tabular}{|l|c|c|c|}
         \hline
         Sample & Tc-yes intrinsic & Tc-maybe intrinsic & Total\\
         \hline 
         \cite{van_eck_henize_2000a} & 29 & 104 & 133\\
         \cite{yang_infrared_2006} & 31 & 255 & 286\\
         \cite{chen_infrared_2019} & 41 & 172 & 213\\
         \cite{uttenthaler_interplay_2019} & 103 & - & 103\\
         \cite{shetye_s_2018,shetye_observational_2019,shetye_s_2021} & 23 & - & 23\\ 
         Total after cross-matching and removing duplicates & 137 & 381 & 518\\ 
         Total after additional cuts & 134 & 378 & 512\\
         Total after ASAS-SN crossmatch \& light curve quality cut & 126 & 314 & 440\\
         Total after parallax error cut & 115 & 243 & 358\\
         Total of final sample with reliable periods measured in this study & 86 & 196 & 282\\
         \hline
    \end{tabular}
\end{table*}

After crossmatching between all samples above, we are left with a total of 518 stars. For most of the stars, we have adopted the Tc detection classifications from \cite{uttenthaler_interplay_2019} as the most up to date, except for the stars in the \citeauthor{shetye_s_2021} papers. We remove some stars that have since been reclassified as Tc-poor based on reevaluations of spectra: these are V530 Lyr and HD 63733 from \cite{shetye_observational_2019}. Interestingly, L$_2$ Pup may also instead be an extrinsic S star, based on new measurements of Tc, which would be consistent with its very low luminosity \citep{uttenthaler_l2pup_2024}.

After collecting tables from these papers, we cross-match to SIMBAD and Gaia with \textsc{astroquery}, and ASAS-SN using the ASAS-SN SkyPatrol python client \textsc{pyasassn}\footnote{The python client takes Gaia DR2 IDs as the input, so we also match DR3 to DR2 IDs. Often the DR3 IDs have been revised from the DR2 IDs.}. We also download and crossmatch the ASAS-SN catalogue of variable stars \citep{christy_asas-sn_2023}, which provides precomputed variability periods and machine-learned variable types. 
To exclude stars with inaccurate luminosities, we employ a cut in fractional parallax error $< 20 \%$ ($\pi/\sigma > 5$) following the recommendations of \cite{andriantsaralaza_distances_2022}. Gaia DR3 parallaxes can often be highly uncertain for AGB stars, because of an uncertain photocentre caused by their large physical size, as well as variability from their intrinsic pulsations and large convective cells. 
Upon crossmatching to ASAS-SN light curve data, we only retain light curves containing n > 300 epochs to select those with ample data, as well as those with mean photometric magnitudes between 8 and 17 to remove light curves with strong saturation or being too dim to have useful light curves. 
Finally, we retrieve dust extinction values A$_V$ from the \cite{lallement_3dmaps_2022} 3D extinction maps using the G-TOMO online `integrated extinction' tool, SIMBAD coordinates and \cite{bailer-jones_estimating_2021} distances.

The spectral types from SIMBAD for the final sample suggest that there are 253 S-type stars, 40 MS-type stars, 34 M-type stars, 15 SC-type stars and 3 C-type stars. The three C-type stars are BH Cru, LX Cyg and V776 Mon - the first two of which have been discussed in prior studies to be SC stars that have recently evolved into C-type \citep{uttenthaler_miraperiods_2011,uttenthaler_lxcyg_2016}. While examining the sample, there was the star HD 111374 (aka. Hen 4-128), which was classified as F-type in SIMBAD. This star was originally in GCSS, and was classified as a possible intrinsic S star in \cite{van_eck_henize_2000b}. We have omitted this star because the luminosity analysis in Section \ref{s:luminosities} places it well below the luminosity expected for intrinsic S stars. We also omit some red supergiants (WOS 7, 34, 39, 49) that were identified as not S-type in \cite{lloydevans_catchpole_sstars_1989}, but classified to be S-type via colour-colour diagrams. 

During the analysis below, we also identify several stars that may be instead extrinsic based on their low luminosities and light curve behaviour, and additional checks for other published characteristics. These stars are GN Lup (Hen 4-155), S1* 513 (Hen 4-171), V2017 Sgr (Hen 4-192) and V957 Cen (Hen 4-134), and will be discussed in more detail in Section \ref{s:lum_comparisons_outliers}. We omit these stars from the final analysis, primarily because they do not fall within the model grids due to their low luminosities.

Additionally, we cross-match our sample with the International Variable Star Index \citep[VSX,][]{watson_vsx_2006} so that we can compare to our pulsation periods measured in Section \ref{s:light_curves}. We were unable to measure reliable pulsation periods for 51 stars during the light curve analysis, so we opted to use the VSX periods for these stars.

A machine learning-based variable type classification from the ASAS-SN g-band catalogue of variable stars \citep{christy_asas-sn_2023}, is available for 221 stars in the sample, 202 of which have a measured variability period. Of these classified stars, there are 132 SRVs, 72 Mira variables, and 17 irregular variables.

Of the 518 stars crossmatched between the above papers, we are left with a final sample of 358 stars with 115 Tc-yes and 243 Tc-maybe stars. The full sample table is available online on CDS\footnote{Refer to online resource Table 2 here, which has names, coords, IDs (Gaia DR3, 2MASS, IRAS, GCSS, ASAS-SN), Tc measured (y/n), references for Tc or intrinsic classification, and SIMBAD spectral type.}. 

\section{METHODS}
\subsection{Light curves and pulsation modes}\label{s:light_curves}

\subsubsection{Light curve analysis}

Light curves for the 358 sample stars were retrieved using the Python client for ASAS-SN SkyPatrol V2.0 \citep{hart_skypatrol2_2023}, which provides epoch photometry of our target stars up to the end of May 2024. 
Initial cleaning of the light curves involved visual inspection of both the V- and g-band light curves; most stars had light curves with both photometric bands available, though a fraction had only g-band photometry.
We only included photometry with the `good' quality labels, and photometric errors $< 0.05$ mag. We also perform sigma-clipping by removing outlying data points that are more than 3$\sigma$ from the median. 
Sometimes, photometry from specific cameras did not appear to behave well, such as cases when data from a single camera stayed at a roughly constant magnitude, or was offset with respect to the other cameras. Mira variables also sometimes had amplitudes that were large enough so that the curves partially fell below the limiting magnitude. 

A prominent issue for the brightest stars in the sample was saturation of the ASAS-SN cameras. Previous data releases of ASAS-SN included corrections to saturated photometry with a method originally from the ASAS survey \citep{pojmanski_asas_2002}. This method involves taking data from the bleed trails around saturated sources, and adding them back into the source photometry as described in \cite{kochanek_all-sky_2017}. Currently, this method appears to be not yet implemented for the newest ASAS-SN photometry, resulting a scatter in photometry for the recent data points of saturated light curves. Many of these are Tc-rich stars, which is consistent with their brighter intrinsic luminosities leading to a higher likelihood of saturating the ASAS-SN cameras.

Next, we computed the Generalised Lomb-Scargle periodogram \citep{lomb_1976,scargle_1982,vanderplas_understanding_2018} for the collection of light curves using the \texttt{LombScargle} module in \texttt{Astropy}. For multiperiodic semi-regular variables, we attempt an iterative whitening procedure, using a version of the code \textsc{balmung}\footnote{\url{github.com/danhey/balmung}}, modified to return false-alarm probabilities for the identified periods. This method first computes the Lomb-Scargle periodogram of a light curve and identifies the highest peak in the periodogram. It computes the Fourier power spectrum density (PSD) unnormalised periodogram, so that the periodogram power is the squared amplitude of the Fourier components at each frequency. Note that we do not include the photometric uncertainty in the time series data so that the PSD interpretation of the periodogram units holds. The frequency and amplitude of this peak is used to fit a sinusoid to the light curve, and this model is subtracted from the light curve data. This process is repeated on the residual light curve in each step, until there are no periodogram peaks above a S/N threshold which we set as 5 times the median of the periodogram power. This results in a list of periods, amplitudes and false-alarm probabilities for the peaks detected in the periodogram. During each step, we also check the phase diagram using the peak periods to visually inspect the reliability of the period. We also inspect the window periodogram to check for any seasonal aliasing effects in the data; this usually reveals peaks at $\sim$ 1 month and $\sim$ 1 year. We apply a cut off in false-alarm probability of $< 10^{-7}$ for a period detection to be considered reliable at each whitening step, and mark any peaks below this as uncertain. 

A total of 45 stars had light curves from which we could not measure and assign a pulsation period. This was either due to saturation issues, a strong long secondary period that made small amplitude variability difficult to constrain, or irregular behaviour. This reduced the sample of stars with ASAS-SN pulsation periods to 282. Approximately half (23/45) of these were Tc-yes stars, which constitutes 21\% of the Tc-yes sample -- these usually saturated the ASAS-SN photometry due to their bright intrinsic luminosities and closer distances. To measure more reliable periods for these stars, we looked to the archival photometry from the Kyoto/Kiso/Kamogata Wide-field Survey (KWS) \citep{maehara_kws_2014}\footnote{\url{http://kws.cetus-net.org/~maehara/VSdata.py}}. This survey has long-baseline ($>10$ year) light curve data for many bright variable stars in the northern hemisphere in the Johnson V-, Ic- and B-bands, which warrant further investigation in the context of long period variable stars. These light curves were also complemented with V-band light curves from the ASAS survey. We were able to make period measurements for 26 of the 45 saturated stars, bringing the total count of stars with no reliable periods down to 19. For the remaining stars, we separately gathered pulsation periods from VSX and assigned them to pulsation modes. Of these, 8 did not have VSX periods, with 3 having no VSX entry at all (BD+79 156, CSS 1230, S1* 628), though they appear to be variables from ASAS-SN data. 
After follow up analysis on the ASAS and KWS light curves, we report revised and new pulsation periods for the following notable examples: CD$-$32$^{\circ}$5117 has a period of 28.8 d, instead of the 180 d period in VSX (a possible half-year alias effect), and TV Aur has a period of 82 d with an LSP of roughly 1000 d instead of the 182 d in VSX. New periods for some stars with no VSX period include: HD 288833 (16.1 d, possible 3rd overtone mode Tc-yes), HD 120179 (44.5 d \& 60.8 d, ASAS 134918-5522.9), NQ Pup (37 \& 51 d, Tc-yes), BD$-$18$^{\circ}$2608 ($\sim60$ d and short LSP, see Section \ref{s:low_mass_cases}) and CY Cyg (148 d, an irregular Tc-yes SC-type star). We also report a new likely LSP of 497 d for the Tc-yes S-type star V1139 Tau. This brought the final number of stars with pulsation periods to 317.

\subsubsection{Pulsation mode assignment}\label{s:mode_assignment}

Although the present sample of intrinsic S stars have variability periods available in published results (including the ASAS-SN g-band catalogue), these only list their dominant variability period. This becomes an issue when attempting to assign pulsation modes, as many SRVs have dominant long secondary periods that are likely not due to radial pulsations (more details in Section \ref{s:lsps}). Furthermore, many SRVs have been shown to pulsate in more than one radial mode \citep{kiss_multiperiodicity_1999}, which necessitates the classification of several variability periods that may be present. 

\begin{figure*}
    \centering
    \includegraphics[width=\linewidth]{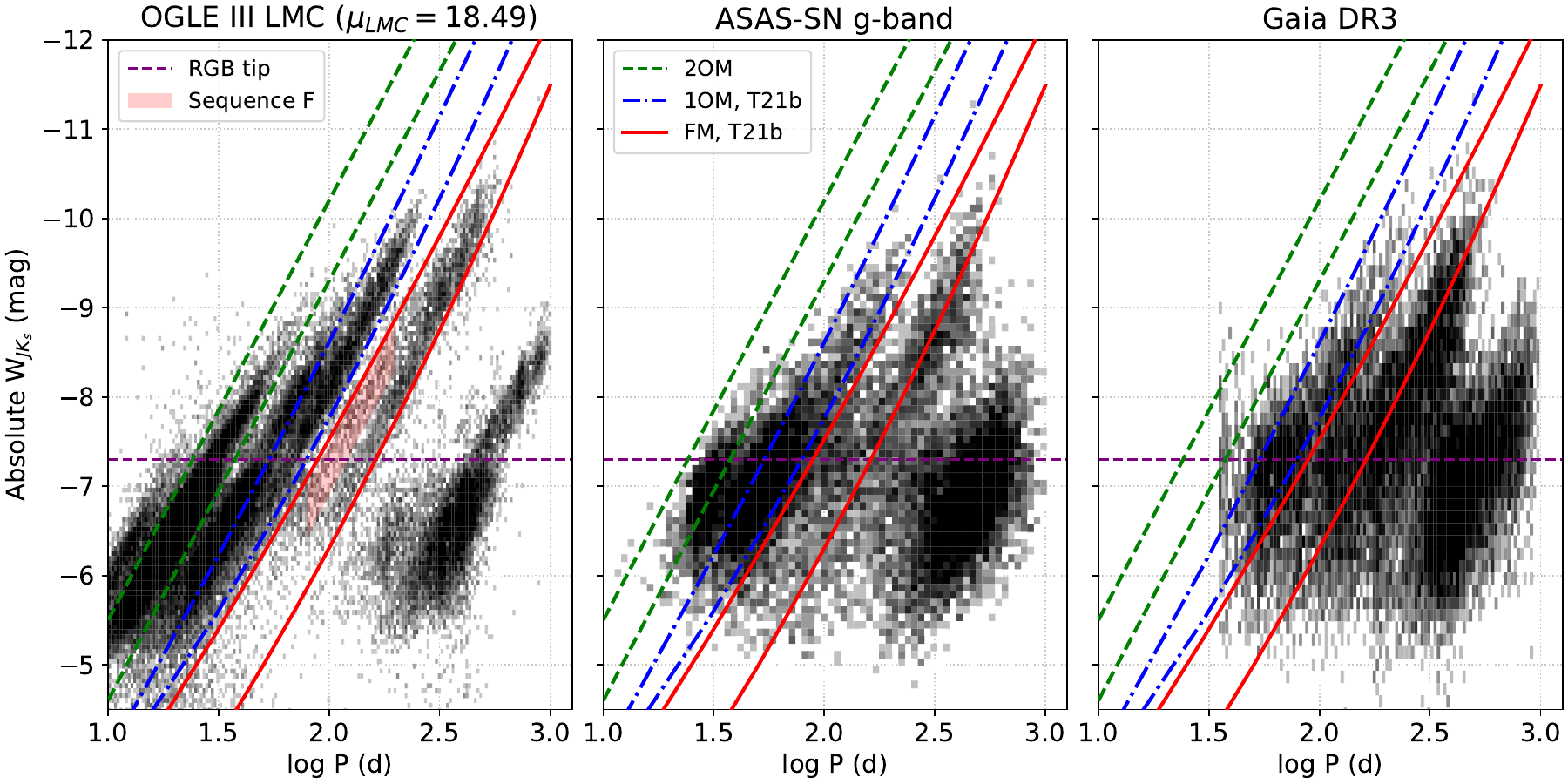}
    \caption{Period-luminosity diagrams for the OGLE III Collection of LPVs in the LMC, the ASAS-SN g-band catalogue of LPVs, and the Gaia DR3 catalogue of LPVs. Lines are from \protect\cite{trabucchi_semi-regular_2021}, and denote the boundaries adopted for sequence C$'$ (first overtone mode, blue dash dot) and sequence C (fundamental mode, red solid) for the LMC PL diagram. We also include lines denoting sequence A (2nd overtone mode, green dashed). We mark the approximate luminosity of the RGB tip (purple dashed line). In the ASAS-SN and Gaia DR3 panels, sequence C appears to be offset with respect to the boundaries; this may be an effect of uncertain Gaia DR3 parallax distances \citep{bailer-jones_estimating_2021} for the more evolved, brighter stars pulsating in the fundamental mode.}
    \label{fig:pldiagrams_ogleasassngdr3}
\end{figure*}

Once the pulsation periods were measured, we then assigned them to a pulsation mode and each star to a variability class. We find that 203 out of 345 ($58\%$) stars can be classified as SRVs, which often display multiple variability periods and require more careful pulsation mode assignment. We employ a mode assignment technique similar to that of \cite{trabucchi_semi-regular_2021}, which assigns radial pulsation modes of stars based on their location in the PL diagram. We retrieved data from the ASAS-SN g-band catalogue of variable stars \citep{christy_asas-sn_2023} and the OGLE III collection of long period variable stars in the LMC \citep{soszynski_optical_2009}, to construct background PL diagrams. For the ASAS-SN data in these diagrams, cuts were made to only include stars with Gaia DR3 fractional parallax errors of $\lesssim 7\%$ $(\pi/\sigma > 15)$, variability amplitudes $> 0.3 $ mag to improve the definition of PL sequences, and to only include stars that were classified as SRVs and Mira variables in the ASAS-SN machine-learning classification pipeline. The OGLE data combines the measurements for the Miras, SRVs and OSARGs. We show the periods from each catalogue against the absolute magnitude calculated from the dereddened Wesenheit index $W_{K_s,J-K_s} = K_s - 0.686 \cdot (J-K_s)$, using 2MASS J- and K$_s$-band photometry. This is converted to an absolute magnitude using the \cite{bailer-jones_estimating_2021} distances (described in Section \ref{s:dists_dust_ext}) for the ASAS-SN catalogue, and a distance modulus of $\mu = 18.49$ for the OGLE LMC variables. This results in recognisable pulsation mode sequences in the PL diagram for each diagram, as shown in Figure \ref{fig:pldiagrams_ogleasassngdr3}. 

We define the PL sequences according to \cite{trabucchi_new_2017}, with stars in sequence A pulsating in the second overtone mode, B and C$'$ in the first overtone mode, and C in the fundamental mode. Periods that lie on sequence D were defined as LSPs; these stars had smaller amplitude variability (higher overtone mode pulsation) along with LSP variability. SRVs often had multiple periods that were assigned to the second overtone, first overtone and fundamental mode, while Miras had single dominant pulsation periods that were assigned to the fundamental mode.

We also find some pulsation periods that may be assigned to the sparsely populated sequence F, between sequences C$'$ and C \citep{wood_pulsation_2015}, which can be seen clearest in the left panel of Figure \ref{fig:pldiagrams_ogleasassngdr3}. These stars usually have an additional period on sequence B or C$'$, and are SRVs. We interpret these pulsation periods as fundamental mode pulsation, as initially suggested in \cite{wood_pulsation_2015}. A possible explanation for this sequence discussed in \cite{trabucchi_semi-regular_2021} is that these stars have switched to dominant fundamental mode pulsation near the end of the TP-AGB interpulse phase, which is predicted by the stellar evolution and pulsation models as shown later in Figure \ref{fig:max_gr_models}. As the star expands along the interpulse phase in earlier thermal pulses, the fundamental mode becomes more unstable and dominant in the final section of the interpulse phase, which may manifest itself in the less luminous sequence F. Contrary to \citeauthor{wood_pulsation_2015}'s explanation that these are high-mass luminous LPVs formed by recent star formation, these stars may instead be lower mass stars, as they spend more time in dominant fundamental mode pulsation along the TP-AGB compared to stars of higher mass. Sequence F would then be less populated because the time spent in dominant FM pulsation near the end of the interpulse is quite short. This fundamental mode period is quickly cut off by a thermal pulse, contraction, and a return to higher overtone mode pulsation (likely back to the 1OM). 

Comparing the OGLE and ASAS-SN PL diagrams, there appears to be an offset in sequence C, where the ASAS-SN sequence (and also our sample stars) lie to the lower right of the LMC sequence. This also may affect another subset of the sample, which appear to show fundamental mode periods beyond sequence C, while having shorter periods that lie near sequence F rather than C$'$. We discuss this offset in Section \ref{s:g2mass_pldiagrams}.

The PL sequences in the Magellanic Clouds show finer structure that has been associated with non-radial oscillations, as the SRVs in the higher overtone mode sequences appear to show stochastically excited oscillations \citep{wood_pulsation_2015,yu_lpv_2020,trabucchi_pastorelli_lpvs_2025}. The same diagram constructed with the ASAS-SN catalogue does not show this fine structure, likely due to the large uncertainties in 2MASS $K_s$-band magnitudes or uncertain distances. Although the results from light curve analysis show signs of non-radial oscillation, in this study we assume pulsations to be solely radial for simplicity. 

\begin{figure}
    \centering
    \includegraphics[width=\linewidth]{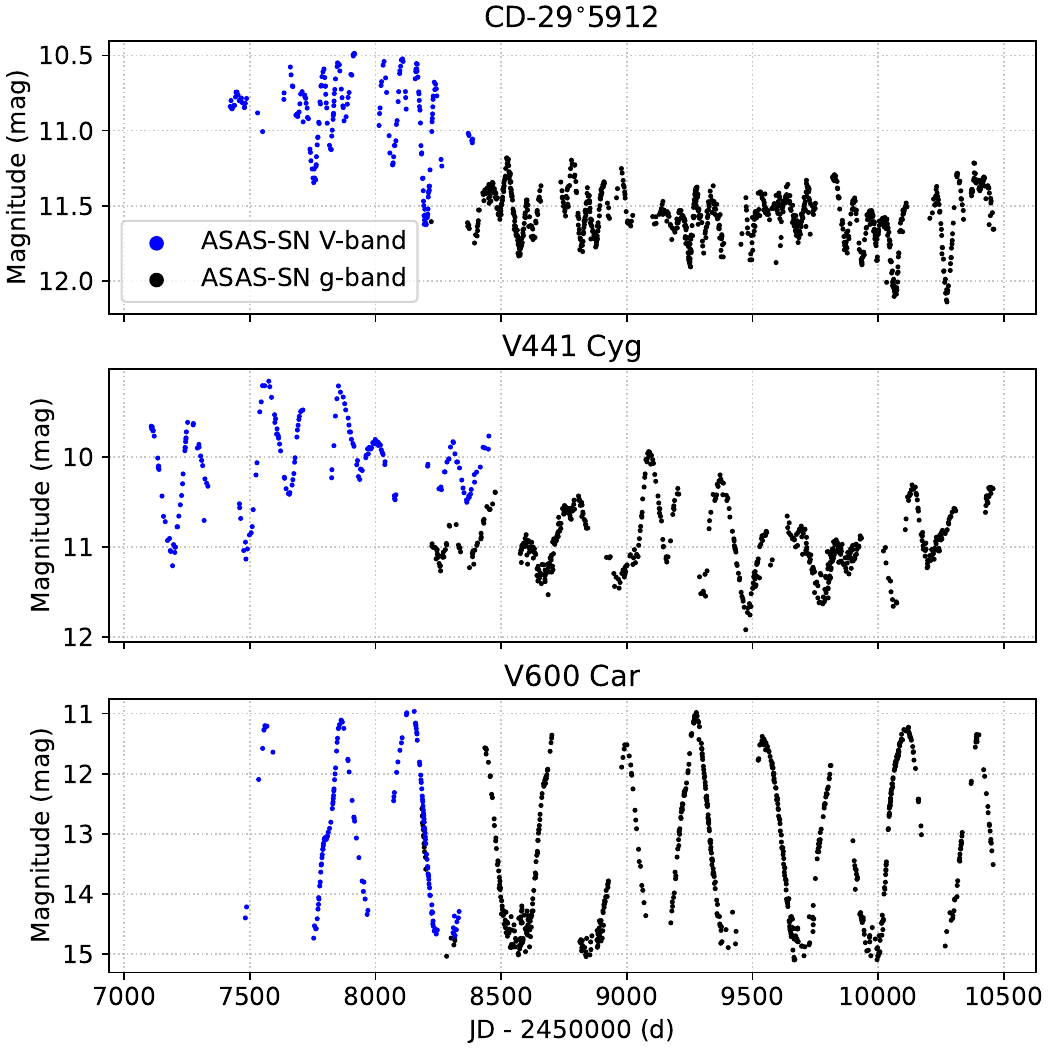}
    \caption{Example ASAS-SN light curves for three Tc-rich S-type stars. Top: Semiregular variable CD$-$29$^{\circ}$5912 (Hen 4-44), middle: SRa variable V441 Cyg (Hen 4-227), bottom: Mira variable V600 Car (Hen 4-109). }
    \label{fig:example_lcs}
\end{figure}

\begin{figure}[t]
    \centering
    \includegraphics[width=\linewidth]{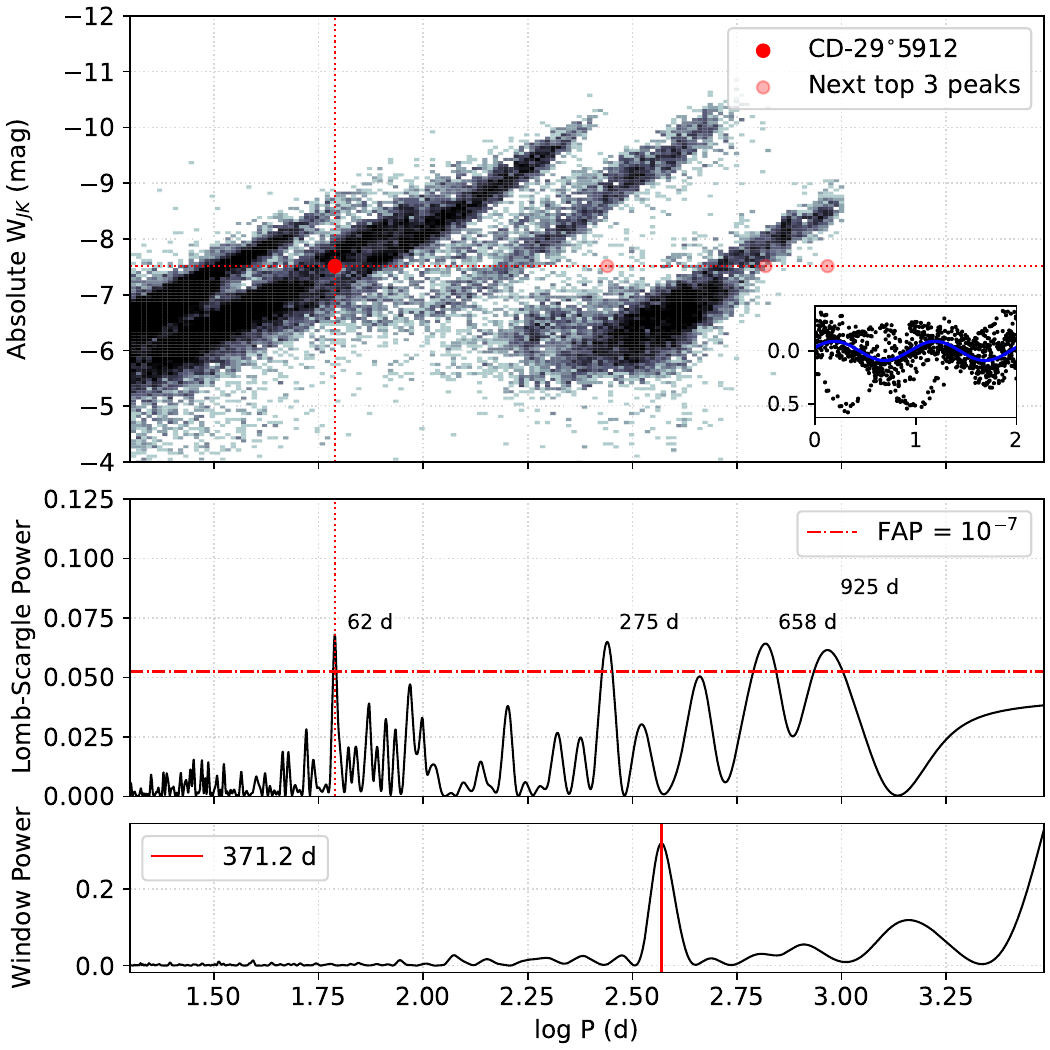}
    \caption{Example light curve analysis and mode assignment for Tc-rich S star CD$-$29$^{\circ}$5912. Top panel: OGLE period-luminosity diagram, and the periods from the periodogram at the absolute W$_{\text{JK}}$ magnitude of CD$-$29$^{\circ}$5912. Top panel inset: Phased light curve showing the 62 d peak period from the periodogram. Middle panel: Lomb-Scargle periodogram of the g-band ASAS-SN light curve. Red horizontal dotted line is the power at which the false alarm probability is 10$^{-7}$, and the peaks above this threshold are labelled with the corresponding period. The 62 d period lies in between pulsation sequences B and C$'$, is thus assigned to the first overtone mode. Bottom panel: Window power spectrum of the light curve, showing structure at $\sim$ 1 yr (371 d).}
    \label{fig:example_lc_analysis}
\end{figure}

\subsection{Luminosities} \label{s:luminosities}

The determination of accurate luminosities for AGB stars is fundamental to estimating their masses. In this section, we evaluate several different approaches to calculating the absolute bolometric luminosities for the sample of intrinsic S-type stars. 

\subsubsection{Distances and dust extinction}\label{s:dists_dust_ext}

To find accurate absolute luminosities, we must obtain accurate distances to the stars in our sample. We use the distances from \cite{bailer-jones_estimating_2021}, who take a probabilistic approach to deriving stellar distances with Gaia EDR3 (early data release) parallaxes. They provide `geometric' distances to objects, which are based on parallaxes, uncertainties and direction-dependent priors on distance, and also `photogeometric' distances which use the colour and apparent magnitude as additional priors. We use the geometric distances as they were available for all of the sample, and both distances were usually consistent to within 250 pc.

The uncertainties surrounding parallaxes for AGB stars is discussed in depth in \cite{andriantsaralaza_distances_2022}. Due to their large physical sizes, large convection cells and often asymmetric structures, AGB stars have uncertain photocentres which introduce significant uncertainty into their parallax measurements. These issues, as well as their high intrinsic brightnesses, mean that the parallax and distance uncertainties of AGB stars are often underestimated \citep{elbadry_gedr3parallax_2021}. We discuss these issues in more detail in Section \ref{s:qual_uncerts}. 

For the above reasons, we introduced a cut in fractional parallax error $< 20 \%$, as mentioned in Section \ref{s:literature_sample}. Roughly 75\% of the sample had both geometric and photogeometric distances (before and after parallax cuts), with both distances agreeing within $\sim$ 10\% up to parallax errors of $<10\%$. Between parallax errors of 10-20\%, the geometric distances may be up to 50\% larger than the photogeometric distances. Distance discrepancies this large seem to affect 13\% of stars that have photogeometric distances. We ultimately opt for the geometric distances because they were available for all stars in the sample, but it is important to note that for some stars these luminosities may be overestimated compared to calculations with photogeometric distances.

Next, we correct for interstellar extinction with the 3D dust extinction maps from \cite{lallement_3dmaps_2022}. We calculate extinction corrections for the near-infrared (NIR) J and K$_s$ magnitudes from the V-band extinction $A_V$ using the relations $A_K = 0.114A_V$ and $A_J = 2.47A_K$, as in \cite{abia_characterisation_2022}. 

\subsubsection{Bolometric corrections based on empirical relations}

We calculate absolute bolometric luminosities using an empirical bolometric correction relation derived by \cite{kerschbaum_bolometric_2010}, hereafter K10. This relation is based on bolometric luminosities derived from SED fitting of a large sample of M- and C-type stars, which is then used to find an empirical relation between the bolometric correction and NIR ($J$-, $K$- and $L'$-band) photometry. We employ this bolometric correction relation as a first check as it is simple to implement using 2MASS $J$- and $K_s$-band photometry, and has precedent in the calculation of luminosity functions of S-type stars \citep{abia_characterisation_2022}. As S-type stars only made up a small proportion of the sample in K10 (31 S stars out of 146 C-stars and 655 M- or K-type stars), they were not analysed in more detail.
However, it is important to note that \cite{kerschbaum_bolometric_2010} identify three groups in the $(J-K)$-$(K-L')$ plane of M-type stars, for which they give three different empirical relations. Upon reconstructing the luminosity function in \cite{abia_characterisation_2022}, we find that they use the `Group A' relation for the intrinsic S stars. Since L'-band photometry is not available for our sample stars, it is difficult to ascertain whether the S-type stars lie in slightly different groups. Ultimately, we also opt for the Group A relation, though it is possible our S stars may still be better described by the other groups. For instance, stars with larger $K-L'$ (Groups B and C) also have larger $J-K$, and so the bolometric correction and therefore the luminosities are underestimated. The Group A box in fact ends at $J-K = 1.6$, so stars with $J-K$ greater than this are likely in a different group. An additional issue here is that the 2MASS $K_s$-band ($2.15$ $\mu$m) is not equivalent to the $K$-band used in K10 ($2.2$ $\mu$m), which would likely induce a systematic offset in these luminosity estimates. 

\citeauthor{abia_characterisation_2022} also note that uncertainties in the absolute $K_s$-band magnitude ($M_{K_s}$) and the absolute bolometric magnitude ($M_{\text{bol}}$) are dominated by distance uncertainties, with an estimated $\pm0.25$ mag as the typical error for the full sample of their stars. We find a similar result for the uncertainty in absolute bolometric luminosity, though our results also include the uncertainty in $A_V$ from \cite{lallement_3dmaps_2022}.

For a luminosity estimate that is independent of Gaia DR3 parallax distances, we also calculate a bolometric luminosity from the PL relation from \cite{andriantsaralaza_distances_2022} for Mira variables, where we use 
\begin{equation}\label{eq:a22_pl}
    M_{\text{bol}} = (-3.31 \pm 0.24)[\log P - 2.5] + (-4.317 \pm 0.060).
\end{equation}
This relation was derived for the O-rich Mira variables in the Milky Way, and is valid for periods between 276 and 514 days. \cite{andriantsaralaza_distances_2022} assume the slope of the PL relation is invariant with chemical type, and use the slope for the PL relation for C-stars in the LMC from \cite{whitelock_agb_2008} to find that the zero-point for the O-rich relation agrees with the C-star relation. Thus we make the assumption that the S-type stars are also reasonably well described by the O-rich Mira group and make luminosity estimates for stars with pulsation periods in the recommended range.

\subsubsection{Bolometric corrections based on SED fitting}

As an alternative to the empirical relation method described above, we also perform fits to the spectral energy distribution (SED) using the online SED fitting tool VOSA \citep{bayo_vosa_2008}. This tool was also used to analyse the SEDs of AGB stars in open clusters by \cite{marigo_fresh_2022}, which includes three S-type stars.

We input the coordinates, \cite{bailer-jones_estimating_2021} geometric distances with uncertainties, and \cite{lallement_3dmaps_2022} extinctions into the VOSA web interface. We retrieve photometry with VOSA, which uses the CDS Vizier database \citep{ochsenbein_vizier_2000} to build an SED. We use photometry from AKARI, APASS DR9, DENIS, Gaia DR3 (Gaia G, BP, RP, and synthetic photometry based on the BP/RP spectra), IRAS, MSX, SDSS, WISE and 2MASS to span a wide range across the visual to infrared. 

We then used VOSA to make a $\chi^2$ fit to the SED using the grid of O-rich GRAMS models, using the default parameter space for the model grid fits. We use the option in VOSA that takes a statistical approach to fitting, where a Gaussian random noise is introduced for 100 virtual SEDs based on the observed SED. The uncertainties of the final parameters are then based on the standard deviation of the values obtained from the 100 SED fits, unless the uncertainty is less than half of the model grid step for the parameter, in which case half the grid step is reported. We have included an example SED fit using VOSA in the appendix (Figure \ref{fig:egvosafit}).

The uncertainty in the bolometric luminosity is obtained by error propagation of the distance and flux uncertainties. Although the uncertainties for these luminosities appear to be quite small compared to those obtained from the empirical relations, the uncertainties for SED model fitting is quite challenging to reliably quantify due to additional complexities that arise from both observations and model atmospheres \citep[see discussion in][]{mcdonald_pyssed_2024}. For the majority of the S-type stars the O-rich models were the best fit, though some stars (the SC-type Miras) were better fit by the C-rich GRAMS models.

\subsubsection{Luminosity comparisons and outliers}\label{s:lum_comparisons_outliers}

\begin{figure*}
    \centering
    \includegraphics[width=\linewidth]{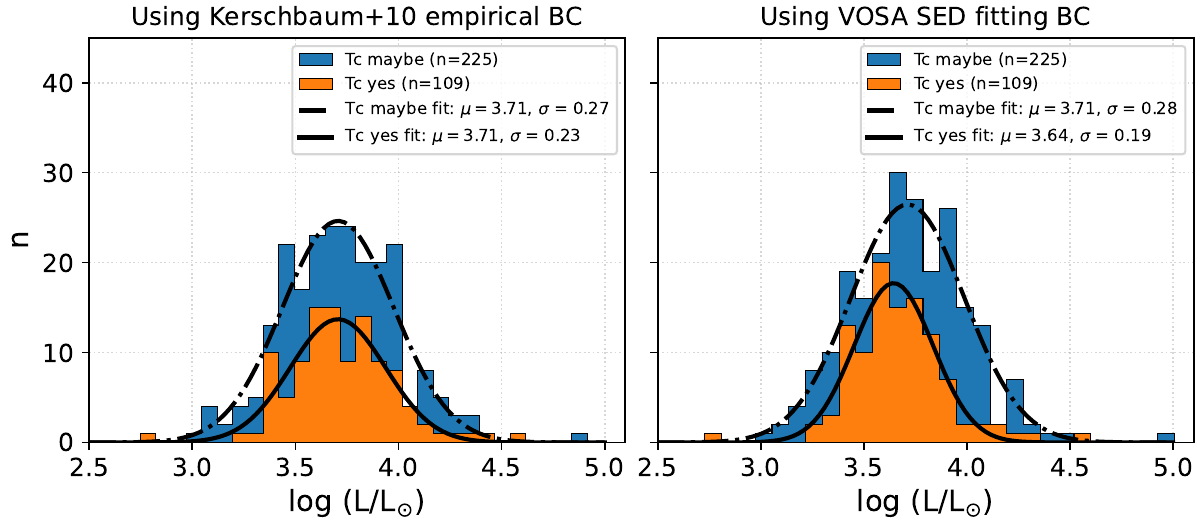}
    \caption{Luminosity functions derived for the sample, using the                             \protect\cite{kerschbaum_bolometric_2010} bolometric corrections (left) and VOSA SED fitting (right).}
    \label{fig:loglsun_k10_v_hists}
\end{figure*}

\begin{figure}
    \centering
    \includegraphics[width=\linewidth]{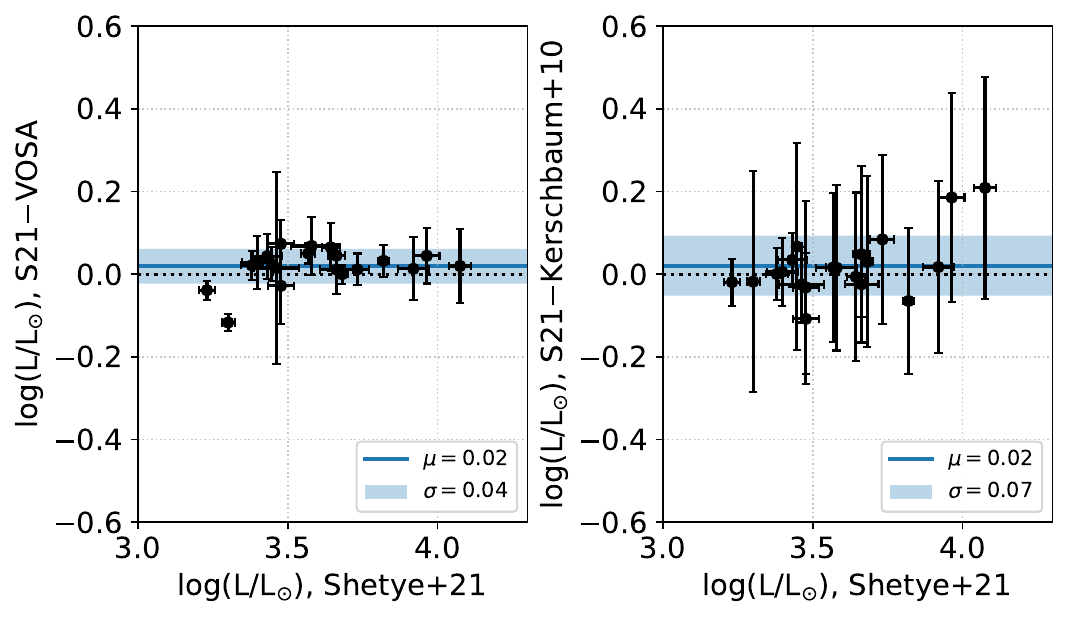}
    \caption{Comparison of residuals between the luminosities derived in this paper and the S21 sample.}
    \label{fig:compare_s21_vosa_k10_lums}
\end{figure}

We now compare the derived luminosities for each method, as well as the distributions for the Tc-yes and Tc-maybe groups. We also compare to the luminosities from S21 to assess the reliability of the derived luminosities. 

In Figure \ref{fig:loglsun_k10_v_hists} we compare the luminosity distributions of the full sample of intrinsic S-type stars based on the bolometric correction method for both Tc-yes and Tc-maybe stars. We fit a normal distribution to each histogram, and find that the mean $\mu$ and standard deviation $\sigma$ are quite consistent between both methods, as well as between the Tc-yes and Tc-maybe groups. The mean of the VOSA luminosities is slightly lower, which may reflect the assumption in the K10 relation that all stars fall into the same group. 

As further verification, we also compare to the luminosities derived for the S21 sample of Tc-rich S stars. These luminosities were derived using high resolution spectra and a grid of MARCS model atmospheres for S-type stars, so are a good independent check for our derived luminosities. We show one to one diagrams of the luminosities in Figure \ref{fig:compare_s21_vosa_k10_lums} for the 20 stars shared with S21. The majority of stars lie within the 1$\sigma$ uncertainties for each method, though there are a few lower luminosity stars from S21 that are less consistent with the VOSA luminosities. We note that the variability type also appears to have an effect on the scatter when comparing the methods as illustrated in Figure \ref{fig:lum_residuals_vartype}, with Mira variables having significantly more scatter. This is likely due to the fact that Miras still have relatively large amplitudes even in the near-IR bands (e.g. \citealt{catchpole_miras_1979}), which would also affect the single-epoch photometry of 2MASS and contribute to differences between the luminosity methods.

We will now highlight a few stars that have notably low luminosities that are significantly below the RGB tip luminosity. Intrinsic TP-AGB stars are not expected below these luminosities, but a few appear to have positive detections of Tc from previously published studies. 

\textit{Hen 4-16}: Identified as Tc-rich in \cite{van_eck_jorissen_tc_1999}, Hen 4-16 is a low-luminosity outlier in the sample of Tc-yes S-type stars. Its luminosity appears to lie well below the commonly accepted luminosity cutoff for TDU of $M_{\text{bol}} \sim -4$ mag, at roughly $\sim -2$ mag \citep{abia_characterisation_2022}. Its parallax distance is $1534 \pm 34$ pc, with only a $\sim2\%$ parallax uncertainty. We find a consistently low luminosity of $612 \pm 78$ \Lsun\ with the K10 empirical relation, and $646 \pm 32$ \Lsun\ with VOSA. It is a small-amplitude variable, with a g-band full amplitude of $\sim0.2$ mag and a measured period of 26.3 d. Interestingly, the periodogram of the V-band light curve has a peak at $18.6$ d, which is further supported by the five available sectors of TESS light curves for this object. On the PL diagram, these periods would place Hen 4-16 in alignment with sequences A and B, for the 18.6 d and 26.3 d periods respectively. Interestingly, it also appears to have a possible additional variability period at roughly 270 d, which lies on the LSP sequence.

However, there is the possibility that Hen 4-16 is not in fact on the TP-AGB, and is actually an RGB star near the tip of the RGB. The lower portions of the PL sequences are also populated with red giant branch variables \citep{ita_trgb_2002,kiss_bedding_rgbpulsation_2003,trabucchi_new_2017}, pulsating in radial order modes. 

The Tc measurement for this object comes from \cite{van_eck_jorissen_tc_1999}, which reports the S/N of the Tc line blends at 4262 and 4238 \AA\ to be 13 and 7 respectively, among the lowest in their sample. A follow-up Tc measurement for this star could resolve this ambiguity, with a negative Tc detection more firmly defining a minimum luminosity for TDU. A positive detection, however, would confirm this star's peculiarity. 
If this star is indeed confirmed to be Tc-rich, a potential explanation could be that it is in the `trough' of the luminosity during the pulse-interpulse cycle. The pulsation models also predict that during a thermal pulse, the overtone modes may once again dominate as a result of the contraction of the stellar radius during either the thermal pulse onset (described as the `knee' region in \citealp{joyce_rhya_2024}), or during the start of the interpulse phase during which the H-burning shell begins to recover its luminosity after being almost extinguished by the thermal pulse. TP-AGB stars near their minimum luminosities in the pulse-interpulse cycle may still be found as overtone mode pulsators, even at late thermal pulses. Despite this possibility, Hen 4-16 lies significantly below the grid of models in the PL diagram, meaning its luminosity is too low to be explained by the TP-AGB. Additionally, the relatively short amount of time spent in the trough of the interpulse luminosity also makes it much less likely for a star to be observed in this phase\footnote{Though, considering the large number of long period variable stars that have been observed, these TP-AGB `trough' stars may exist in the data. See the discussion on HD 288833 in Section \ref{s:low_mass_cases}}. Whether this is the case for Hen 4-16 depends on its status as a Tc-rich star, for which we recommend follow-up Tc measurements. For the remainder of the paper, we exclude Hen 4-16 from analysis on the count of its ambiguous classification, as well at it not being consistent with the grid of stellar models we explore.

\textit{GN Lup (Hen 4-155) \& S1*513 (Hen 4-171)}: These stars are both classified as intrinsic in \cite{van_eck_henize_2000a}, based on their multivariate clustering method. The light curves for both stars appear to show regular periods of 320 and 397 d respectively, which are in the range expected for Miras. Both are also noted as outliers in the IRAS-2MASS colour-colour diagrams of \cite{van_eck_grid_2017}. The authors remark that these objects have very large $K-[12]$ colour excesses, and are perhaps more appropriately classified as SC-type. This large excess in the longer wavelengths is also reflected in the SED, with a poor fit in VOSA at longer wavelengths even when trying both O-rich and C-rich grids. A further look into these stars with SIMBAD and VSX show that GN Lup is a visual binary, and is noted as such in VSX. Interestingly, S1*513 has a foreground star within 2``, identified in Gaia DR3 -- this is within the ASAS-SN pixel resolution of 7.8``.

\textit{V2017 Sgr (Hen 4-192, Plaut 3-1347)}: V2017 Sgr has been classified as a Tc-rich Mira variable, with a period between 409-426 d \citep{uttenthaler_tc_2007,iwanek_ogle_2022}. We measure a period of 406 d, which is reasonably close to, but on the lower side of published values. A notable feature of its light curve in ASAS-SN is its alternating peak brightness (perhaps from a higher order pulsation mode), with the minimum of the light curve appearing to be below the limiting magnitude. The Gaia DR3 distance for this star is $615^{+65}_{-63}$ pc, which leads to luminosities of L$_{\odot, VOSA} = 57.5 \pm 12.9$, L$_{\odot, K10} = 84.1 \pm 20.5$. This conflicts with \cite{uttenthaler_tc_2007}, where V2017 Sgr is assumed to be a bulge star (distance of $\sim$8 kpc), to find a luminosity of $\sim 14,000$ \Lsun. Using the \cite{andriantsaralaza_distances_2022} PL relation with the 406 d period gives L$_{\odot, A22} = 5847 \pm 1332$, which also deviates from the above luminosities. It seems likely that the distance is severely underestimated, likely related to the fact that this star is located towards the bulge. 

\textit{L$_2$ Pup}: A low-luminosity SRV star, the K10 and VOSA methods estimate discrepant luminosities of 1432 $\pm$ 465 \Lsun\ and 452 $\pm$ 50 \Lsun\ respectively. Both are consistently lower than the RGB tip luminosity, although the SED was not very well fit with VOSA. Furthermore, this star's position on the Gaia-2MASS diagram is also rather odd, lying in the low-luminosity `C-rich' region of the diagram despite being an M-type star. This star has previously been classified as an AGB star \citep[e.g.][]{kervella_l2pup_2014}, as well as a SRV showing pulsations with a period of 140 d that appear to show both Mira-like coherent and stochastic excitation mechanisms \citep{bedding_l2pup2_2005}. \cite{cunha_stochastic_2020}, however, classify L$_2$ Pup as a classical pulsator with a coherent driving mechanism. During writing, \cite{uttenthaler_l2pup_2024} made updated Tc measurements for this star that reclassify it as Tc-poor instead of Tc-rich, and also identify it as a fundamental mode pulsator based on its position in the PL diagram. The same paper attributes the large $W_{JK}$ index to significant circumstellar extinction, which may be caused by carbon chemistry from the disc structure around this star \citep{kervella_l2pup_2014}. However, models of RGB pulsation in \cite{trabucchi_new_2017} only extend to $\sim 100$ d, though the pulsation of L$_2$ Pup may be complicated by its chemistry, disc-structure and potential close companion. We have chosen to exclude this star from the following analysis as its status as a true intrinsic AGB star is uncertain.

\subsection{Grids of stellar models}

\subsubsection{Detailed AGB stellar models}

We make use of the detailed TP-AGB models from \cite{karakas_stellar_2016}, supplemented with a higher resolution grid of initial masses between 0.9 to 1.5 \Msun, as described in Table \ref{tab:tpagb_grid}. The grid covers three values for the initial global metallicity Z (0.014, 0.007, 0.0028), with canonical helium abundances Y of (0.28, 0.26, 0.25) respectively. We assume that solar metallicity is Z = 0.014 \citep{asplund_2009}, and the other metallicities are assumed to have a scaled-solar composition. The lowest initial mass model of 0.9 \Msun\ is intended to serve as a low limit of realistic initial masses, given the age of the Universe of 13.8 Gyr, and that the time taken to reach the TP-AGB for a star of this mass would be $\sim18$ Gyr at Z = 0.014, or $\sim 14.6$ Gyr at Z = 0.007. Due to convergence issues for the 2.0 \Msun\ model for Z = 0.007, we use a 2.1 \Msun\ model instead. We include a Reimers' mass-loss rate on the red giant branch with $\eta = 0.4$ for masses up to and including 1.5 \Msun, as the much longer RGB lifetimes for these lower masses would cause more envelope mass to be lost relative to the total mass. On the AGB phase, we use the \cite{vassiliadis_wood_1993} mass-loss prescription for all models. 

\begin{table}
    \centering
    \begin{tabular}{|c|c|}
        \hline
        Initial mass M$_{\text{i}}$ (\Msun) & 0.9, 1, 1.1, 1.2, 1.3, 1.4, 1.5, \\ & 1.75, 2*, 2.5, 3, 3.5, 4, 5 \\
        \hline
        Initial metallicity Z$_{\text{i}}$ & 0.0028, 0.007, 0.014 \\
        \hline
    \end{tabular}
    \caption{Grid of TP-AGB models. \\
    *: Z = 0.007 uses 2.1 instead of 2.0.}
    \label{tab:tpagb_grid}
\end{table}

\subsubsection{Radial pulsation models}

We combine the above detailed stellar evolution models with the grid of linear pulsation models constructed in T19. These models are based on a code by \cite{wood_strange_2014} \citep[see also:][]{fox_theoretical_1982,keller_cepheids_2006}. The model constructs a spherically symmetric, static envelope, and conducts a linear stability analysis of its structure. They apply this to a relevant grid coverage with TP-AGB models computed with the COLIBRI stellar evolution code \citep{marigo_evolution_2013,marigo_parseccolibri_2017}. 

T19 also include a convenient interpolation code (\textsc{giraffe}) to find pulsation characteristics given a set of global stellar parameters acquired from detailed stellar evolution codes. These input parameters are the luminosity log(L/\Lsun), the effective temperature log(T$_{\text{eff}}$), total mass, core mass, C/O ratio, and the elemental mass fractions of hydrogen, helium and metals (X, Y, Z). The code calculates pulsation periods for the radial modes from the fundamental to the fourth overtone mode, as well as the amplitude growth rate, which acts as a proxy for the pulsation amplitude which cannot be predicted in linear models. The growth rate \footnote{$GR_n = \exp \left( 2 \pi \frac{\omega_{\text{R},n}}{\omega_{\text{I},n}}\right) - 1$, where $\omega = \omega_{\text{R}} + \textbf{i}\omega_{\text{I}}$ is the eigenfrequency from the time dependence of the perturbations of form $\xi \propto \exp(\omega t)$. See T19 for more details.} 
describes the `fractional rate of change in the radial amplitude per pulsation cycle', and provides a general indication of the dominant pulsation mode, though it is subject to uncertain interactions between convection and pulsation. It is important to note that for SRVs, the observed amplitude of a given mode is likely to vary over time as a result of stochastic excitation \citep[e.g.][]{christensendalsgaard_srvsolarlike_2001}. This may make the growth rate less reliable as a proxy for observability, though it should remain a good indicator of which modes we expect to be excited for a given time in the models.
The models account for the surface chemical enrichment during the TP-AGB, which can alter opacities, stellar radi and pulsation properties. More specifically, the grid explores surface C/O ratios. It is however important to note that their results actually show that the effect of varying C/O makes minimal difference to pulsation parameters when C/O < 0.95 (i.e. while the opacity has a minimal effect), which should be the case for this sample of S-type stars. This advantage is of course useful for completeness when exploring the full TP-AGB in models, but is not expected to significantly affect most of our target stars. The grid also covers three values of the mixing-length parameter \alphamlt\ (1.5, 2.0, 2.5). This allows for flexibility for comparisons to different stellar codes and choices of \alphamlt.

From comparisons of their models to PL sequences in the Magellanic Clouds, T19 found good agreement for the overtone modes but over-predict fundamental mode pulsation periods. This discrepancy has been identified before \citep[eg.][]{lebzelter_wood_47tuc_2005}, and is mainly due to the rearrangement in stellar structure due to high amplitude pulsation in the fundamental mode. This result motivated a follow-up paper \citep{trabucchi_modelling_2021}, which uses a 1D hydrodynamic code to explore the non-linear regime of large-amplitude pulsations. These models appear to better fit observational fundamental mode PL sequences in the Magellanic Clouds, finding that the non-linear periods are systematically shorter than the linear periods. In order to calculate expected pulsation periods using these results, the study also includes analytical period-mass-radius relations based on best fits to their models. We also use this best-fitting period-mass-radius relation to calculate a `non-linear' fundamental mode period to compare with our measurements. However, this does not take into account any direct dependence on metallicity or C/O. While a detailed study of how nonlinear pulsation is affected by changing composition is yet to be explored, \cite{trabucchi_modelling_2021} find qualitative changes similar to the linear case, where fundamental mode periods in C-rich models are reduced by 10-15\% and the onset of dominant fundamental mode pulsation is delayed to higher luminosities and larger radii. How composition affects this pulsation, especially approaching C/O $\sim 1$, requires further investigation.

\subsubsection{Pulsation period evolution during the TP-AGB}\label{s:pulsation_models_tpagb}

To calculate expected pulsation periods and growth rates for each radial mode over the TP-AGB, we input the stellar parameters from detailed model results of the Monash stellar evolution code into the \textsc{giraffe} interpolation program by T19. We also calculate the fundamental mode periods in the non-linear regime using the best-fitting period-mass-radius relation from \cite{trabucchi_modelling_2021}, and assume that the growth rates for these periods are similar to those in the linear regime. Specifically, we compute these pulsation characteristics: a) at the maximum luminosity of each interpulse, just before the onset of a thermal pulse, and b) for every 10th model number for the TP-AGB model results.

\begin{figure}[t]
    \centering
    \includegraphics[width=\linewidth]{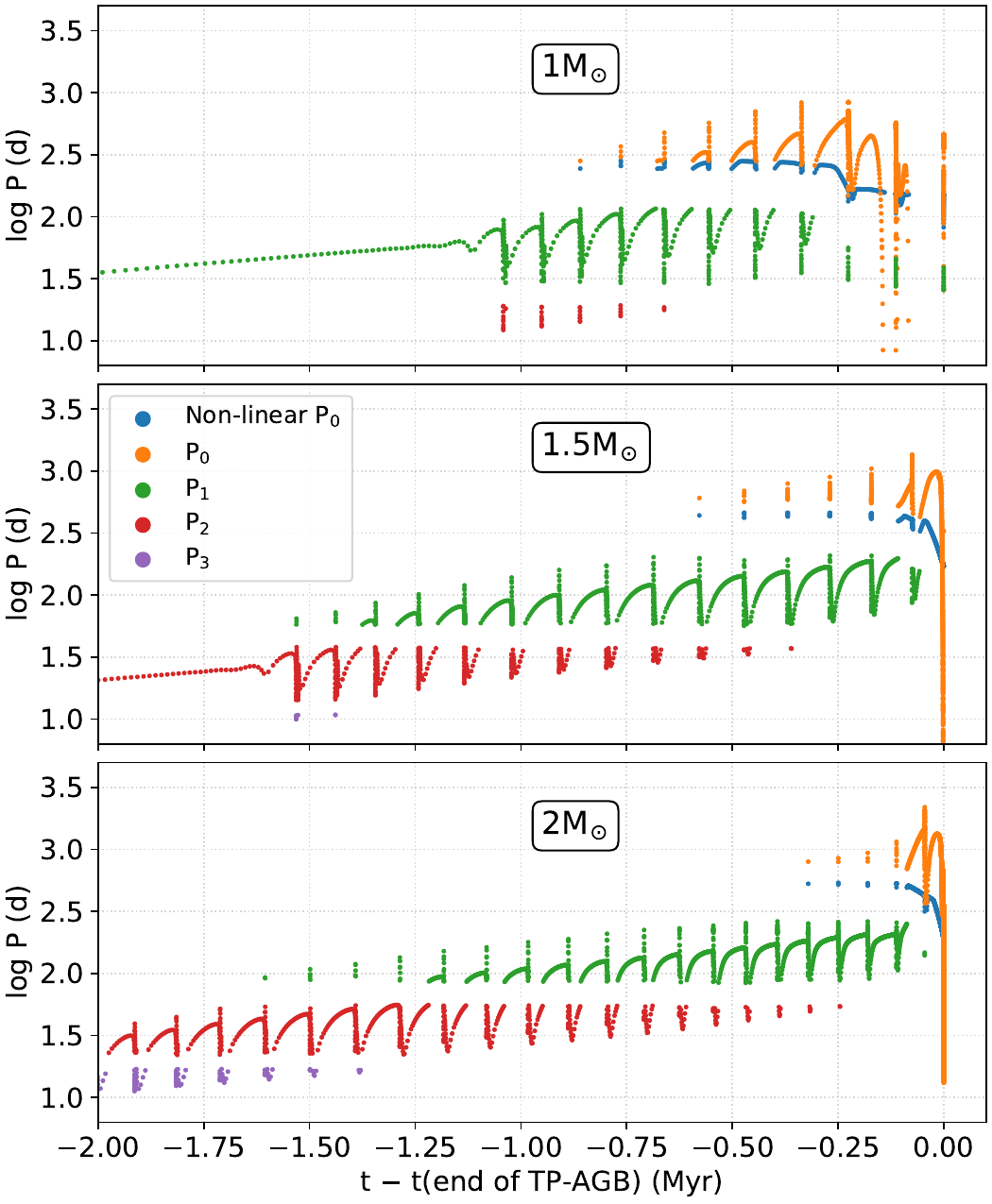}
    \caption{Theoretical pulsation periods against time during the TP-AGB for three solar metallicity (Z = 0.014) models of initial mass 1, 1.5 and 2 \Msun, combining the Monash stellar evolution models and \protect\cite{trabucchi_modelling_2019,trabucchi_modelling_2021} pulsation models. This includes the pulsation periods for the third, second and first overtone modes, as well as the linear and non-linear fundamental mode period. Only the pulsation mode with the largest amplitude growth rate is shown for each model time, to illustrate the expected dominant pulsation mode throughout evolution. We show both the linear and non-linear fundamental mode periods at the same time to demonstrate the difference. Note that the sharp dips and peaks at each thermal pulse are overrepresented in this diagram, because we only include every 10th model, and the model time steps shorten during thermal pulses.}
    \label{fig:max_gr_models}
\end{figure}

An interesting prediction of the T19 models is that the dominant mode of radial pulsation evolves over the pulse-interpulse cycle, according to the growth rates. In Figure \ref{fig:max_gr_models}, we shown the period of pulsation mode with the largest growth rate against time for models with initial mass 1, 1.5 and 2 \Msun\ with solar metallicity (Z = 0.014). The proportion of time spent in a dominant pulsation mode throughout the TP-AGB is markedly different with increasing stellar mass -- a 1 \Msun\ star is expected to be a dominant fundamental mode pulsator (Mira) for a much longer fraction of its TP-AGB lifetime than a 1.5 \Msun\ star. We thus expect TP-AGB stars of lower masses to be more likely observed pulsating in lower order pulsation modes, and vice versa for higher masses. This also means that TP-AGB stars near the end of their lives (in their final few thermal pulse cycles) may still be observed as SRVs.

\section{RESULTS}\label{s:results}

\subsection{Pulsation periods and modes}\label{results_periods_modes}

\begin{figure*}
    \centering
    \includegraphics[width=\linewidth]{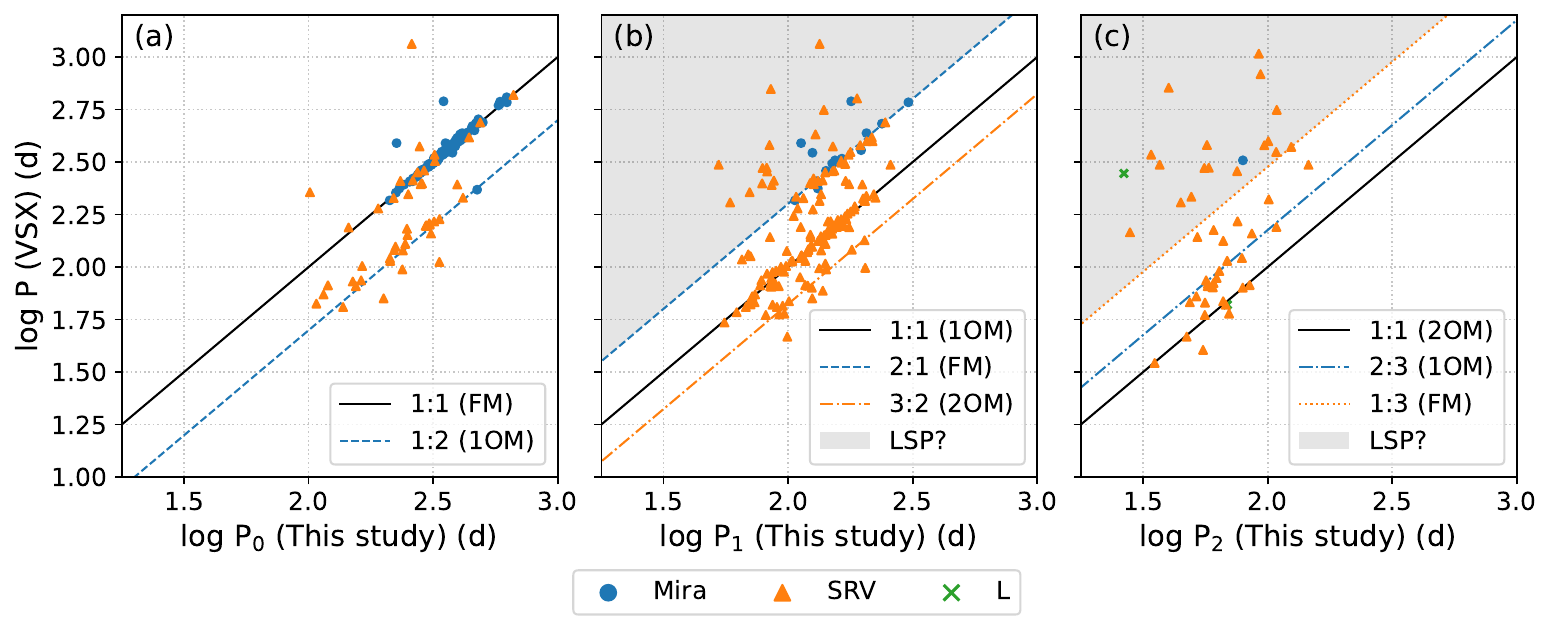}
    \caption{Comparison of pulsation periods measured in this study with VSX periods. Each panel is a one to one diagram comparing the assigned pulsation mode period with the star's VSX period, where (a) is the fundamental mode, (b) is the first overtone mode and (c) is the second overtone mode. Miras are denoted as blue circles, SRVs as orange triangles and the irregular variables (L) as green crosses. The majority of our fundamental mode periods agree well with VSX values, while the overtone mode periods illustrate a variety of measured periods that can be associated with different pulsation modes, as well as dominant long secondary periods. We also include lines for the 2:1 and 3:2 period ratios roughly expected for separate harmonics of each assigned mode. The shaded region in panels (b) and (c) is a domain where the VSX period is much longer than our measured period, and is likely a dominant long secondary period. Further details of these trends are discussed in Section \ref{results_periods_modes}.
    }\label{fig:comparevsx_measured_per}
\end{figure*}

The periods listed in the published results for SRV stars often only include the dominant period, while these stars show more complex behaviour caused by multiple variability periods. We make a consistency check between the pulsation periods we measure for our sample and periods from existing measurements. Figure \ref{fig:comparevsx_measured_per} shows one to one diagrams of the our measured periods against the periods from VSX, for each assigned pulsation mode. We also include lines of expected period ratios to demonstrate that studies can often pick up the periods of other pulsation modes (or LSPs) for SRVs. 

Most fundamental mode periods in panel (a) of Figure \ref{fig:comparevsx_measured_per} appear consistent with published values -- the majority of these stars on the one to one line are large amplitude Miras with clearly defined periods. These fundamental mode periods mostly lie between 180 to 600 d. There also are a number of SRVs with both fundamental and first overtone mode periods, but with more scatter. Most of these include SRa variables that have a dominant first overtone mode (measured in VSX), but also have fundamental mode periods that we measure in this study.

There are a few periods that are somewhat longer in existing measurements than the periods measured in this work. These are WRAY 18-94, WRAY 18-165, GQ Peg and V441 Cyg (Tc-rich). WRAY-18-94 appears to be a SRV with an erroneous long period (615 d, compared to our measured 348 d) after inspection of the ASAS light curve. Similarly, WRAY 18-165 has a listed period of 1155 d which may be a LSP, but we measure a clearer 259 d that better reflects the observed variability (longest VSX period in panel (a) of Figure \ref{fig:comparevsx_measured_per}). GQ Peg has a clear 225 d period from ASAS-SN light curves, which is closer to the 248 d period identified in NSVS \citep{wozniak_nsvs_2004} compared to the 367 d period in ASAS. Finally, we measure a period of 279 d for the Tc-rich SRa variable V441 Cyg, which agrees with the period in \cite{uttenthaler_pmasslossmiras_2013}, but not with the GCVS period of 375 d.  

Panels (b) and (c) of Figure \ref{fig:comparevsx_measured_per} compare the measured first and second overtone mode periods to the VSX periods respectively. Both show significantly more scatter, which may be explained by the measurement of several simultaneously excited pulsation modes, as well as the presence of long secondary periods in many of these stars. Many of first overtone mode periods show agreement between our values and VSX values, with over-densities corresponding to the fundamental mode and possibly the second overtone mode. There is significantly more scatter for the second overtone modes in panel (c) -- a few periods clearly agree well, with some others likely corresponding to the first overtone mode. There are longer periods that can be associated with long secondary periods that have much larger amplitudes than the pulsations, and are measured as the primary period in VSX. Another source of this scatter may come from differences in available data quality, as we have found that the periods reported for SRVs appear to be inaccurate after following up on the light curves. Often these are stars with periods from the automated period search from the ASAS-3 survey \citep{pojmanski_asas_2002}, which may be subject to the effects of window aliasing or outliers (for example, CD $-$32$^{\circ}$5117 from Section \ref{s:light_curves}). While studies such as \cite{vogt_miras_2016} have made follow up measurements for Miras using ASAS-3 data and compared them to VSX values, it appears that this is yet to be investigated for the SRVs.

\subsection{Gaia-2MASS and period-luminosity diagram}\label{s:g2mass_pldiagrams}

\begin{figure*}
    \centering
    \includegraphics[width=\linewidth]{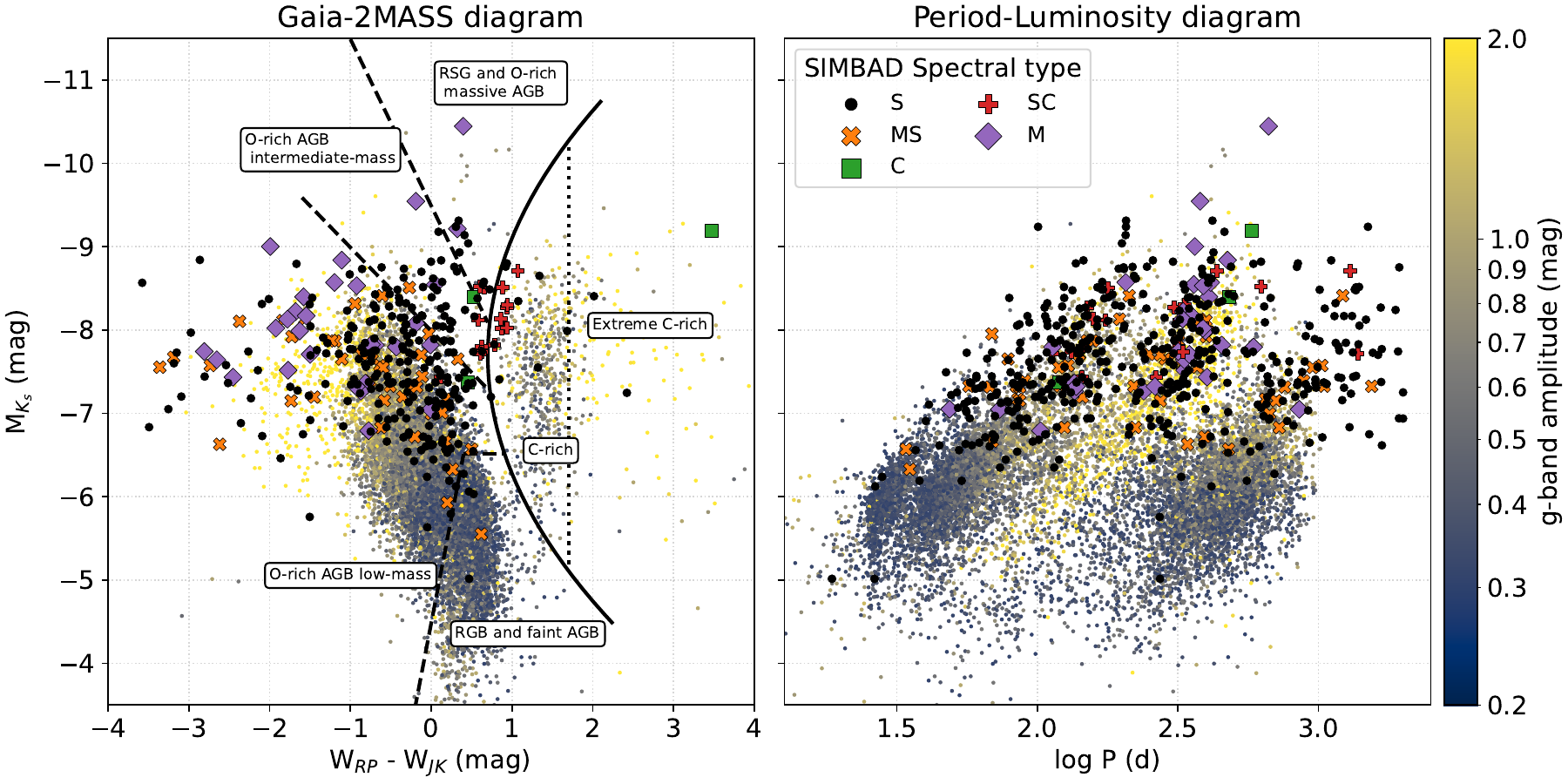}
    \caption{Left: Gaia-2MASS diagram of the sample, overlaid on the ASAS-SN g-band catalogue of LPVs. Markers and colours for the sample are spectral types from SIMBAD. The regions in the Gaia-2MASS diagram are from \protect\cite{lebzelter_new_2018} but shifted to absolute magnitudes using $\mu = 18.49$, and indicate groups of initial stellar mass and current surface chemistry. Right: Period-luminosity diagram, with the same data as the left panel. Periods for the sample are from this study, and periods for the ASAS-SN catalogue are as published in the catalogue. Both panels use the absolute K$_s$-band magnitude.}
    \label{fig:g2m_pl_sample_simplespt}
\end{figure*}

\begin{figure*}
    \centering
    \includegraphics[width=\linewidth]{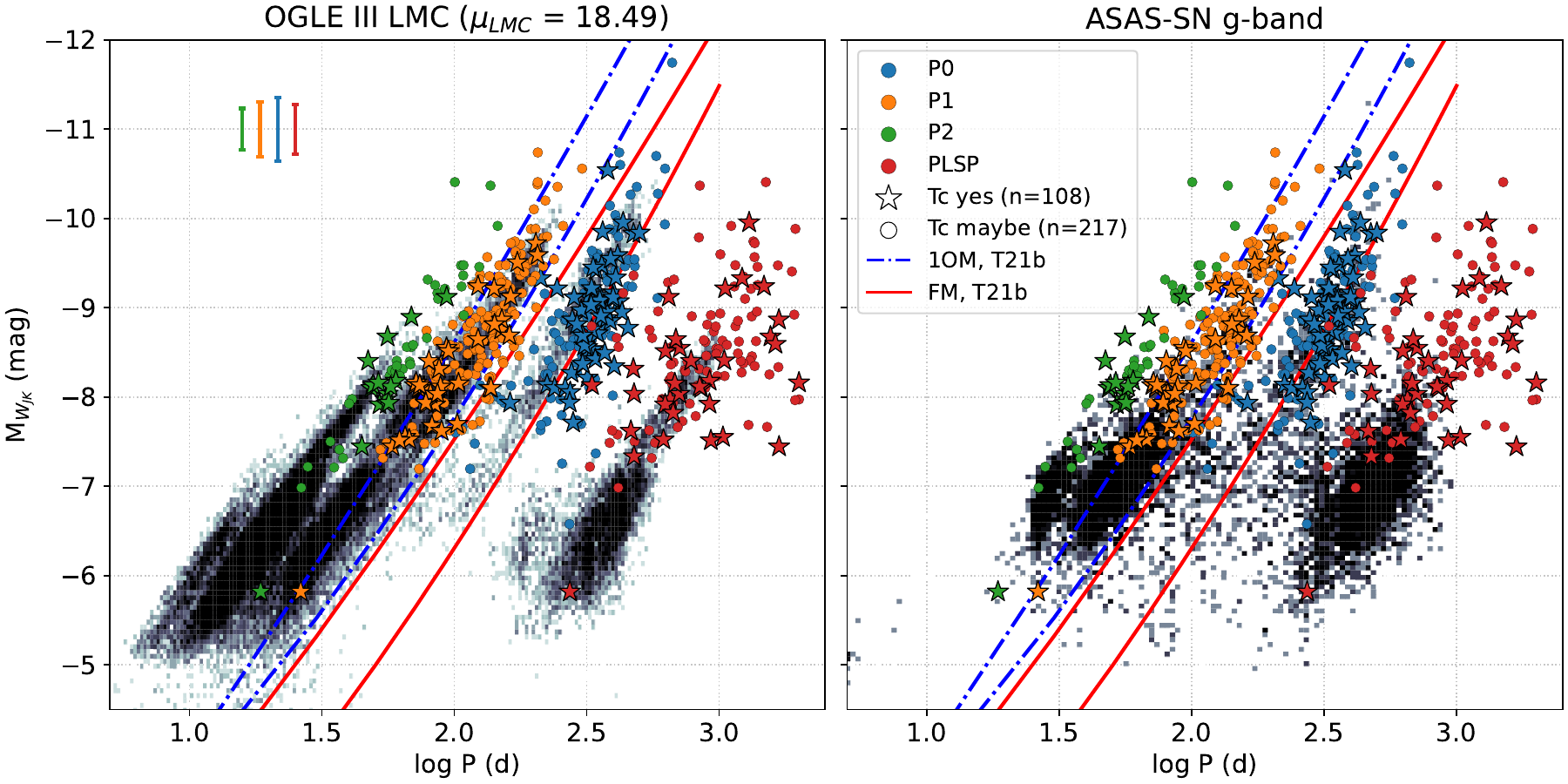}
    \caption{Period-luminosity diagrams of LPVs in the OGLE III collection for the LMC (left) and the ASAS-SN g-band catalogue (right) compared to the sample of intrinsic S-type AGB stars with reliable periods measured in this study (large symbols). We show all determined periods for each star, including those with multiple periods, which appear at the same luminosity. Boundaries for the LMC pulsation sequences are from \protect\cite{trabucchi_semi-regular_2021} as in Figure \ref{fig:pldiagrams_ogleasassngdr3}. Average error bars for stars with P2, P1, P0 and PLSP (left to right) are also shown in the left panel. Both panels use the absolute Wesenheit index magnitude $MW_{JK}$.}
    \label{fig:pl_allsurveys_sample}
\end{figure*}

A useful tool to diagnose the evolutionary state of the sample of S-type stars is via the side by side Gaia-2MASS \citep{lebzelter_new_2018} and PL diagrams in Figure \ref{fig:g2m_pl_sample_simplespt}. For the catalogue data in these diagrams, we also colour the points by their g-band variability amplitude as measured in the ASAS-SN catalogue. The inclusion of the variability amplitudes better defines the PL sequences, and also acts as an additional proxy for the evolutionary progression of these stars, with higher amplitude stars generally being more evolved (Miras). We begin by describing the general expected behaviour of a long-period variable TP-AGB star in both of these diagrams. 

The Gaia-2MASS diagram (Figure \ref{fig:g2m_pl_sample_simplespt}, left panel) separates the O-rich and C-rich populations of red giant stars based on the difference in their Wesenheit index calculated from visual and NIR photometric colours ($W_{RP}-W_{JK}$), where $W_{JK}$ is as defined in Section \ref{s:mode_assignment} for the 2MASS $J$- and $K_s$-bands and $W_{RP}$ is calculated from Gaia RP and BP photometry with $W_{\text{RP}} = G_{\text{RP}} - 1.3 \cdot (G_{\text{BP}} - G_{\text{RP}})$. This discrimination is based on how the chemistry affects the BP$-$RP colours of these stars; O-rich stars have brighter BP$-$RP colours, with an almost constant J$-$K$_s$ value, so move to the left of the diagram. On the other hand, the BP$-$RP for C-rich stars become dominated by molecular bands, while J$-$K$_s$ is only mildly affected, so that W$_{JK}$ dominates and the star moves to the right of the diagram. The mass and chemistry regions of this diagram are based on results of population synthesis with the \textsc{TRILEGAL} code \citep{girardi_trilegal_2005} for stars on the AGB. The qualitative behaviour in which the more evolved stars move further to the left or right side of the diagram is supported by the gradient in variability amplitude, as more evolved stars develop larger amplitude pulsation. A curious feature on the C-rich side is the existence of some lower luminosity stars classified as SRVs, and a follow up of the light curves confirms this. These stars also have lower visual variability amplitudes for a given period compared to SRVs on the O-rich side, which is somewhat expected from the behaviour in Miras where the lower sensitivity of the absorption bands of C-bearing molecules in the optical range compared to O-bearing molecules \citep[e.g.][]{soszynski_optical_2009}. It is unclear if these are the result of intrinsic enrichment (they are SRVs that have undergone sufficient TDU to become C-rich) or extrinsic enrichment, and this would require further investigation. There a clear split between the O-rich and C-rich groups, a result of the rapid transition from O-rich to C-rich composition that these stars undergo and the effect of that transition on the structure of the stellar atmosphere. In the PL diagram (Figure \ref{fig:g2m_pl_sample_simplespt}, right panel), we expect an AGB star to generally evolve toward the upper right, increasing in luminosity while moving across the PL sequences. After a thermal pulse, the star would shift down to shorter pulsation periods, back to sequences associated with higher order pulsation modes.

Now that we have established the general behaviour of TP-AGB stars in both of these diagrams, we can examine where the S-type stars lie. The results are shown in Figure \ref{fig:g2m_pl_sample_simplespt}, where the sample stars are overlaid on the ASAS-SN data as a comparison to LPVs in general. We also include the spectral type classifications from SIMBAD. In the Gaia-2MASS diagram, the majority of S-type stars lie within the the O-rich, low-mass region, spreading out toward the left of the diagram. However, interestingly, there are a significant number of stars within the O-rich to C-rich `gap'. These stars potentially have been `caught' during their switch toward the C-rich side of the Gaia-2MASS diagram, with higher C/O ratios. Notably, the SC-type stars are clustered around $W_{RP} - W_{JK} \sim 1$, in the intermediate region between the O-rich and C-rich groups. 

Next, we focus on the PL diagram for our stars, shown in Figure \ref{fig:pl_allsurveys_sample}. In these panels, we compare the derived pulsation periods and assigned pulsation modes for the sample with the OGLE III LMC sample (left panel) and the ASAS-SN sample (right panel). An important feature to consider for the comparison to the OGLE PL diagram is a clear offset to the lower right (see Figure \ref{fig:pldiagrams_ogleasassngdr3}) that is especially prominent for the fundamental mode sequence, while the overtone modes appear somewhat consistent. This offset may be caused by less accurate parallax distances for the more evolved AGB stars, due to their higher intrinsic luminosities and more extended circumstellar environments from mass-loss. We also expect larger scatter in the 2MASS photometry for nearby bright stars, which likely affect many of the sequence C$'$ and C stars which are intrinsically bright. Additionally, there is evidence that Galactic Miras are fainter than LMC Miras \citep{sanders_miraplgdr3_2023}, which is consistent with this offset. It is also possible that the offset is caused by the difference in metallicity between the two populations. The universality of PL relations of LPVs in different populations of age and metallicity is a point of contention in prior work concerning Miras \citep[][and references therein]{andriantsaralaza_distances_2022}; further investigation in disentangling the effects of distance uncertainties and metallicity will be important when considering Galactic samples of LPVs in general, including other spectral types. There is increasing evidence that the PL sequences for LPVs are in fact affected by chemical composition (Z, X and C/O): either indirectly by affecting the stellar structure through radiative opacity \citep{trabucchi_modelling_2019}, or by affecting the pulsational stability of the envelope \citep{trabucchi_pastorelli_lpvs_2025}. If we assume that the PL sequences are universal even for differing metallicity environments, it is possible that we have underestimated the luminosity of these fundamental mode pulsators, which presents difficulties in estimating masses, as we find in Section \ref{s:fm_mass_ests}. 

\subsection{Initial mass estimates for intrinsic S-type stars}\label{s:mass_ests}

\begin{figure}
    \centering
    \includegraphics[width=\linewidth]{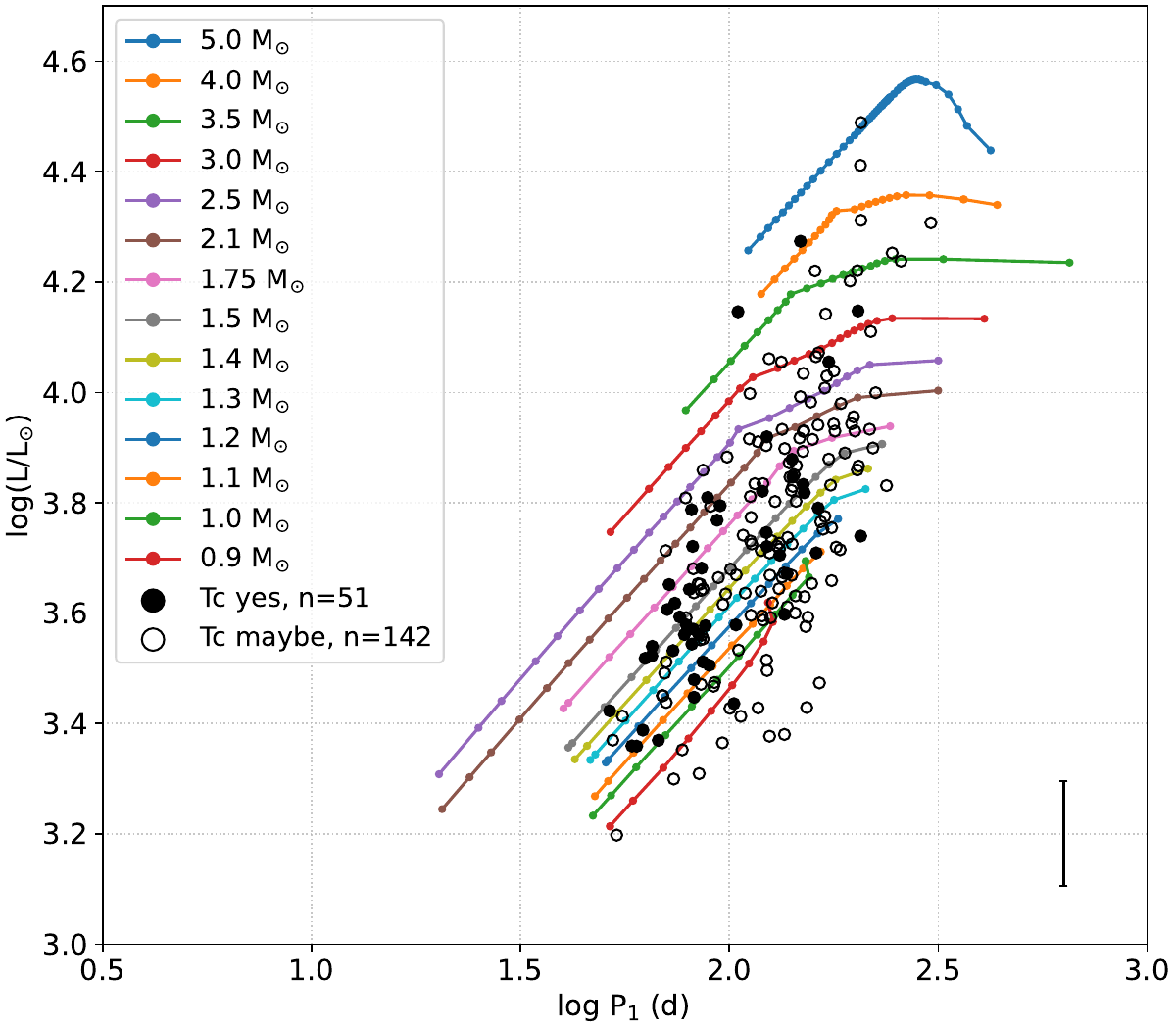}
    \caption{Theoretical period-luminosity diagrams for the first overtone mode with initial model metallicity Z = 0.007, compared with observations. Each point on the model tracks represents the period and luminosity at each thermal pulse, as defined in the text. Filled black circles are Tc-yes stars, open black circles are Tc-maybe and the error bar in the bottom right is the mean luminosity uncertainty.}
    \label{fig:modelpldiagrams_z007_p1}
\end{figure}

\subsubsection{Overtone modes}\label{s:overtone_masses}

\begin{figure}
    \centering
    \includegraphics[width=\linewidth]{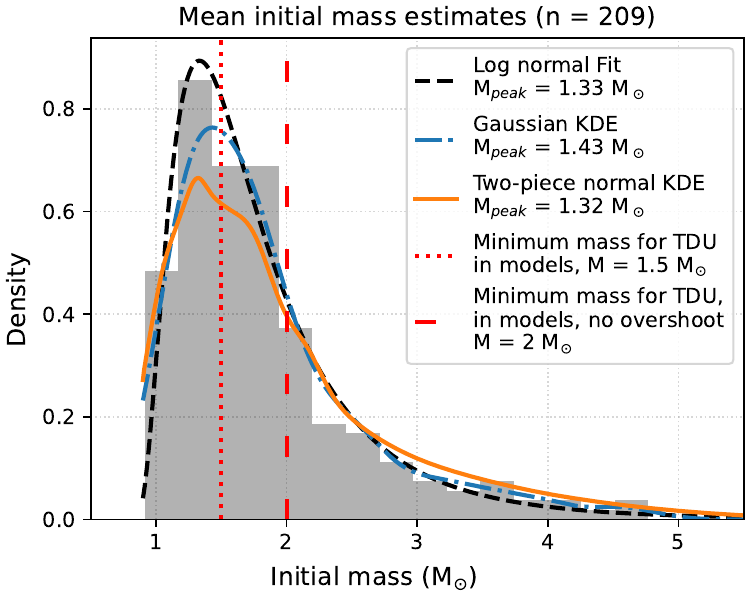}
    \caption{Histogram and distributions for mass estimates using the first and second overtone mode periods, and the mean mass between results from the Z = 0.007 and 0.014 model grids. We also report the peak mass of each distribution according to a log normal fit (dashed), a Gaussian kernel density estimation (dash dot) and a two-piece normal KDE (solid). The Gaussian KDE uses the Silverman formula to estimate the bandwidth, while the two-piece normal KDE uses the upper and lower mass uncertainties as $\sigma$ for each side of the kernel. Vertical lines denote current model estimates for the minimum mass required for TDU (dotted, loosely dashed).}
    \label{fig:mass_ests_mean_masses}
\end{figure}

In order to estimate pulsation masses for our sample, we interpolate between model tracks in the PL plane. As shown in Figure \ref{fig:modelpldiagrams_z007_p1}, the TP-AGB tracks trace arcs in the PL plane for each model mass. We first make cubic polynomial fits to the detailed PL tracks, for each mass and metallicity in the grid. Next, we perform a grid interpolation between these model fits using \texttt{scipy.interpolate.LinearNDInterpolator} and \break \texttt{scipy.spatial.KDTree}, and estimate the mass of a star from this interpolation.

In the present models, the majority of mass loss occurs toward the end of the AGB, when the star is experiencing dominant fundamental mode pulsation. We work under the assumption that earlier on the AGB, while the second and first overtone modes are dominant, the mass loss is small. The linear pulsation models for the overtone modes are well-behaved in the PL plane for the majority of the time, tracing parallel lines of mass until significant mass loss sets in. This behaviour is likely due to the \cite{vassiliadis_wood_1993} mass-loss prescription used in the models, which we discuss in Section \ref{s:mass_z_alpha_massloss}.

We also derive mass uncertainties based on the positions of the upper and lower bounds of the luminosity uncertainties. We neglect the uncertainty in pulsation period, as it is ill-defined from the periodogram analysis (this only yields a false-alarm probability instead of practical uncertainty bounds), and we assume that the period is well-determined such that the mass uncertainties remain within those determined with just the luminosity. The PL tracks are clearly defined for the overtone mode periods, but comparisons of observations and models becomes challenging for the fundamental mode - we discuss these in the next sections. 

In Figure \ref{fig:mass_ests_mean_masses} we visualise the distributions of these estimated masses, with a histogram and a kernel density estimation (KDE). For stars with both second and first overtone mode periods, we estimate a mass by taking the mean of the masses derived from both modes. We find that the 2OM mass distribution contains higher masses on average than for 1OM masses (see Appendix Figure \ref{fig:mass_ests_all}). Another point of note here is that for stars with both periods, the modes may be in fact non-radial; this would lead to slight offsets in the derived masses, which we describe further in Section \ref{s:qual_uncerts}. The final mass estimate is calculated by taking the mean of the mass estimates from the Z = 0.007 and 0.014 model grids, and mass uncertainties are added in quadrature. The mass uncertainties are asymmetric, so we adopt the two-piece normal distribution \citep{wallis_twopiece_2014} as the kernel in one of the KDEs. This distribution uses the asymmetric uncertainties $\sigma_1$ and $\sigma_2$ as the 1$\sigma$ value for the normal distribution on each side of the mode $\mu$, and rescales each side to give a common value at $\mu$. We exclude data points outside of the interpolated region, and set lower and upper uncertainties beyond the boundaries to be the minimum or maximum mass in the grid respectively. To compare with this kernel, we also tested a log-normal fit to the distribution, as well as a KDE which uses a Gaussian kernel with the bandwidth calculated using the Silverman formula \citep{silverman_1986}. 
\newline
Finally, we note that two of the stars were above the luminosities of this model grid: the RSGs V386 Cep (Tc-maybe) and NO Aur (Tc-yes). This is consistent with these stars being variable type SRc, and they appear to have initial masses $> 5$\Msun. V386 Cep was identified as being intrinsic S-type in \cite{guandalini_infrared_2008} and \cite{chen_infrared_2019}, so this object requires follow up Tc measurements.

We find that the peak of the mass distribution for the full sample is roughly 1.43 \Msun\ from the Gaussian KDE, using the masses after averaging over model metallicity and pulsation mode. When isolating the Tc-maybe stars this peak is roughly consistent, but for the Tc-yes stars the histogram peaks at 1.54 \Msun\ when using the Gaussian KDE. However, if we account for the mass uncertainties using the two-piece normal KDE the peak for the Tc-yes stars drops to 1.32 \Msun\ (see Appendix Figure \ref{fig:massest_tcytcm}). 

\subsubsection{Fundamental mode}\label{s:fm_mass_ests}

For the Mira variables pulsating dominantly in the fundamental mode, initial mass estimates become more challenging when considering both the models and observations. Once the fundamental mode becomes dominant in the models, mass-loss according to the VW93 prescription has become significant enough to shift the stellar tracks in pulsation period due to the changing envelope mass. Also, the change in envelope composition due to TDU begins to affect pulsation, as models above 1.5 \Msun\ become carbon-rich, which lengthens the pulsation period via the increased photospheric radii caused by the change in envelope composition. Furthermore, we have the linear and non-linear fundamental mode periods from T19 and \cite{trabucchi_modelling_2021} to compare; the latter was found to more accurately match sequence C in the Magellanic Clouds so we aim to use these. 

The Miras are also where the uncertainties in luminosity likely become more significant. While pulsation periods are clearly defined, due to their intrinsic brightness and more extended envelopes at this evolutionary stage, Miras are likely to have more uncertain parallax distances. Additionally, their larger photometric amplitudes may have stronger effects on the luminosity estimated from both empirical relations and SED fitting. An alternative here is to compare with luminosities from a PL relation, for which we use the relation from \cite{andriantsaralaza_distances_2022}. \cite{sanders_miraplgdr3_2023} derive PL relations for O-rich Mira variables using Gaia DR3 parallaxes, and found that the parallax uncertainties are underestimated for $G \lesssim 11$ mag and $\varpi > 0.5$. While this was only a small fraction of the stars in that paper, 68 of the 76 S-type Mira variables in our sample fall under this criteria; of these, 17 of the 25 Tc-yes stars are likely affected. 

These combined modelling and observational uncertainties for Mira variables make it challenging to approach their mass estimates in the same manner as for the overtone mode pulsators. Upon comparing the observations with models using the non-linear fundamental mode periods, we find that many of the data lie below the tracks, though there is large scatter (see Figure \ref{fig:fm_model_a22_sample}). This is a similar inconsistency to that seen when comparing the OGLE LMC and ASAS-SN/Gaia DR3 PL diagrams as discussed in Section \ref{s:g2mass_pldiagrams}. Interestingly, the \cite{abia_characterisation_2022} PL relation also seems to lie somewhat below the model tracks, a discrepancy which needs further investigation. For the above reasons, in this paper we decide not to report mass estimates using fundamental mode pulsation, though we intend to revisit this issue in upcoming work.

\subsubsection{Effects of $Z$, \alphamlt, mass-loss and pulsation modes}\label{s:mass_z_alpha_massloss}

\begin{figure}
    \centering
    \includegraphics[width=\linewidth]{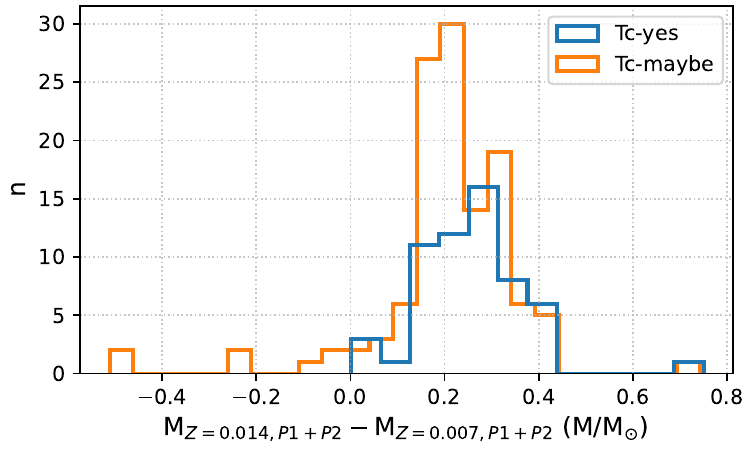}
    \caption{Histogram of the difference between initial mass estimates using models of Z = 0.007 and 0.014. In general, the mass estimates increase with increasing model metallicity Z.}
    \label{fig:mest_z_compare}
\end{figure}

The initial masses derived in this study depend significantly on the model parameters and treatment of physics along the evolution of the TP-AGB. Our model grid covers three initial metallicities (Z = 0.014, 0.007, 0.0028) to cover a reasonable range expected for a Galactic sample. In Figure \ref{fig:mest_z_compare} we illustrate the shift in the mass distributions for model metallicities of Z = 0.014 and Z = 0.007 in the case of the overtone mode pulsators. A decrease in the initial model metallicity from Z = 0.014 to Z = 0.007 shifts the distribution down in mass by roughly 0.2 \Msun, as the model tracks move to higher luminosities and shorter periods. A key caveat in the derived masses is indeed the lack of a metallicity estimate for the majority of the stars, and the assumption of a uniform metallicity for each mass distribution. The metallicity distribution of the S21 sample ranges from [Fe/H] $\approx-$0.70 to solar, which provides some affirmation that the model grid covers an appropriate range of initial metallicities to describe these stars. 

As an additional check of the metallicity effect on the initial mass distribution, we randomly assign a [Fe/H] value to each of the stars in the sample. This random [Fe/H] is drawn from a Gaussian fit to the [Fe/H] distribution for the S21 sample, which gave a distribution with $\mu  = -0.34$ and $\sigma = 0.16$.  We then make a mass distribution that uses the closest model metallicity to the randomly assigned [Fe/H], where we choose Z = 0.0028 for [Fe/H] $\leq -0.5$, Z = 0.007 for $-0.5 <$ [Fe/H] $< -0.2$ and Z = 0.014 for [Fe/H] $> -0.2$. We make these draws 100 times to find the mean Gaussian KDE. The result favours the peak in the Z = 0.007 mass distribution, which is expected given the [Fe/H] distribution has $\mu  = -0.34$. We find this mass distribution to peak at 1.37 \Msun, which is consistent with the result for the average masses in Section \ref{s:overtone_masses}.

The mixing length parameter \alphamlt\ is also a significant uncertainty in detailed stellar evolution models. This parameter can be thought of as a measure of convective efficiency, and an increase in \alphamlt\ generally results in a decrease in effective temperature in stellar models \citep[see a recent review by][]{joyce_mlt_2023}. Although the models we use assume that the value of \alphamlt\ is constant, this is likely unrealistic on the AGB \citep{lebzelterwood_agbvar_2007}. T19 found that the value of \alphamlt\ had some effects on the pulsational stability and periods of their models, with higher values corresponding to slightly larger growth rates for 1OM and FM pulsation, as well as mild effects on pulsation period with only a roughly 5\% increase in the fundamental mode period from a doubling of \alphamlt. 

Treatment of mass-loss during both the RGB and the TP-AGB is also a key component in deriving accurate masses. Mass-loss along the RGB is dependent on the efficiency coefficient $\eta$, which continues to be constrained by studies of globular and open clusters \citep[eg.][]{miglio_ocrgbs_2012,mcdonald_zijlstra_rgb_2015,howell_seismomassloss_2022}. In our stellar models, mass-loss of order $0.1$ \Msun\ occurs on the RGB, which is a relatively small effect compared to the luminosity uncertainties as shown in Figure \ref{fig:modelpldiagrams_z007_p1}.
Pulsation and mass-loss along the TP-AGB are inextricably tied when considering a pulsation-enhanced, dust-driven outflow scenario \citep[see][]{hofner_mass_2018}. Our sample is comprised mainly of semiregular and Mira variables with large amplitude variations, and these stars usually show strong mass-loss. A closer investigation of the mass-loss rates for these stars (eg. connections to IR-excess and dust mass-loss rates such as \citealp{ramstedt_massloss_2009} and \citealp{mcdonald_irhip_2012}), and using different mass-loss prescriptions in our models may give rise to different mass estimates. However, we expect that these mass-loss effects will become most significant for Miras with dominant fundamental mode pulsation, and less so for SRVs. 

Although an in-depth discussion on observed mass-loss rates for our sample stars is beyond the current scope, there are studies such as \cite{ramstedt_sagbmassloss_2006} that have measured mass-loss rates for S-type AGB stars. These observations may assist in making more accurate mass estimates, especially for the stars with high mass-loss rates.

The mass-loss formula of \cite{vassiliadis_wood_1993} used in our models has a dependence on the pulsation period, following the formula:

\begin{equation*}
    \log \dot{M}(M_{\odot} \text{yr}^{-1}) = -11.4 + 0.0125 P (\text{days}),
\end{equation*}

where the fundamental mode period $P$ is calculated from the period-mass-radius relation:

\begin{equation*}
    \log P (\text{days}) = -2.07 + 1.94 \log R/R_{\odot} - 0.9 \log M/M_{\odot}
\end{equation*}

where radius $R$ is the average stellar radius.

This is based on the linear non-adiabatic pulsation calculations for Miras with $0.6 \lesssim M/M_{\odot} \lesssim 1.5$. If this linear fundamental mode period is much longer than observed periods as established in T19 and \cite{trabucchi_modelling_2021}, then the model mass-loss rates may be somewhat overestimated. Recent 3D models of fundamental mode pulsation \citep{ahmad_3dpulsation_2023,beguin_agbparams_2024} also appear to support the idea that fundamental mode periods are shorter than previously established in linear models. This would be an additional effect to consider in detailed models for future attempts to make mass estimates for Mira variables.

Finally, we address the bias in mass that the observed pulsation mode may be imparting on our mass distribution. As mentioned previously (and demonstrated in Figure \ref{fig:max_gr_models}), lower mass AGB stars are more likely to be observed pulsating in lower order radial pulsation modes. Or put another way, for larger initial masses the higher overtone modes remain dominant for longer on the TP-AGB. This may be the reason that we find the 2OM masses to be higher on average than the 1OM masses. However, this observational bias also means that AGB stars with lower initial masses are more likely to be observed as Mira variables, which we have not been able to estimate reasonable masses for. This means that there may actually be more intrinsic S-type stars with lower initial masses, which could act to further decrease the minimum mass for TDU.

\subsubsection{Lowest mass intrinsic S-type stars: interesting cases} \label{s:low_mass_cases}

\begin{table}
\begin{tabular}{|l|l|l|l|l|l|l|l|l|l|l|l|l|}
\hline
Name & Luminosity (\Lsun)  & P1 (d) & M$_\text{init}$ ($M_{\odot}$)\\ \hline 
BD $+$34$^{\circ}$1698* & 3248 $\pm$ 527 & 86.4 & 1.27 $^{+0.26}_{-0.29}$ \\ \hline 
BD $-$18$^{\circ}$2608 & 2285 $\pm$ 183 & 60 & 1.20 $^{+0.14}_{-0.14}$ \\ \hline 
CD $-$29$^{\circ}$5912 & 2443 $\pm$ 307 & 62.1 & 1.25 $^{+0.19}_{-0.22}$ \\ \hline 
CD $-$30$^{\circ}$8296 & 2287 $\pm$ 171  & 58.5 & 1.23 $^{+0.13}_{-0.13}$ \\ \hline 
CSS 182 & 2342 $\pm$ 345 & 67.6 & 1.11 $^{+0.27}_{-0.18}$ \\ \hline 
CSS 454 & 2803 $\pm$ 1496 & 82.3 & 1.14 $^{+1.06}_{-0.36}$ \\ \hline
S Ori & 5491 $\pm$ 588 & 82.3 & 1.14 $^{+1.06}_{-0.36}$ \\ \hline 
TX Peg & 3960 $\pm$ 763 & 135.6 & 1.15 $^{+0.24}_{-0.34}$ \\ \hline
UW Peg & 2728 $\pm$ 446 & 102.6 & 1.00 $^{+0.17}_{-0.11}$ \\ \hline
V Boo & 4694 $\pm$ 429 & 137.5 & 1.25 $^{+0.14}_{-0.17}$ \\ \hline
V335 Sge & 3015 $\pm$ 276 & 82.4 & 1.22 $^{+0.15}_{-0.16}$ \\ \hline 
V812 Oph & 3202 $\pm$ 676 & 89.5 & 1.21 $^{+0.35}_{-0.32}$ \\ \hline 
V915 Aql & 2618 $\pm$ 125 & 80.3 (VSX) & 1.07 $^{+0.08}_{-0.07}$ \\ \hline 
VX Cen & 5119 $\pm$ 271 & 161.4 & 1.21 $^{+0.10}_{-0.22}$ \\ \hline 
Vy 12 (CSS 1152) & 3790 $\pm$ 509 & 103.7 & 1.28 $^{+0.20}_{-0.24}$ \\ \hline 
W Cyg & 5070 $\pm$ 1062 & 131.7 (VSX) & 1.28 $^{+0.20}_{-0.24}$ \\ \hline 
\end{tabular}
\caption{Details of Tc-rich S-type stars with low estimated initial masses. We list their luminosities, first overtone mode periods and initial mass estimates averaged across the results for Z = 0.014 and 0.007. *: Bitrinsic S-type star candidate in S21, see section \ref{s:lsps}. 
}\label{tab:lowest_mass_ests}
\end{table}

In this section, we highlight some intrinsic S-type stars that appear to have initial masses below the $\sim 1.40$ \Msun\ peak. These stars seem to be well below the minimum mass for TDU predicted by detailed models of $\sim 1.50$ \Msun, which make them objects of particular interest for placing constraints on this mass limit. For each, we briefly discuss their pulsation and variability characteristics, their luminosities, and any relevant prior work. In Table \ref{tab:lowest_mass_ests} we also list the Tc-rich stars with initial mass estimates $< 1.40$ \Msun.

\textit{V915 Aql}: This star has been identified as Tc-rich \citep{little_tc_1987,shetye_s_2018}, and most recently was estimated to have an initial mass of 1 \Msun\ in S21. This star is unfortunately saturated in the ASAS-SN g-band, but has a measured variability period of 80.3 d in VSX, originally from the GCVS catalogue \citep{samus_gcvs_2017}.
This variability period can be verified using light curves from ASAS, INTEGRAL-OMC and MASCARA \citep[][see \ref{s:appendix_v915aql} for these light curves]{pojmanski_asas_2002,alfonso-garz_integralomc_2012,burggraaff_mascara_2018}. We compute the Lomb-Scargle periodogram on these light curves to find periods of 79.8 d, 81 d and 82.8 d respectively. V915 Aql was also observed for one sector by TESS, and though the light curve does not show a full period, there is clear variability with a period longer than the $\sim 26$ d sector.
V915 Aql's luminosity also appears to be somewhat inconsistent across both luminosities from this study and what was derived in \cite{shetye_s_2018}. The luminosity derived with the K10 empirical relation places the star at $2085 \pm 1280$ \Lsun, while VOSA lifts this slightly within the K10 luminosity uncertainties to be $2618 \pm 126$ \Lsun. This is brighter than the S21 value of $2000 \pm 100$ \Lsun. 
Placing this star on the PL diagram, it seems to lie on the first overtone mode sequence. Comparing this to models of first overtone mode pulsation periods, the K10 and S18 luminosity places this star significantly below the 1 \Msun\ line for its estimated metallicity (S21, [Fe/H] $\sim -0.5$), while the VOSA luminosity appears to be more consistent with the 1 \Msun\ estimate of S21. Our final mass estimate of $\sim1.1$ \Msun\ is based on the Z = 0.07 model grid. A comparison to the initial mass derivations of S21 is presented in the next section. This star appears to be one of the lowest luminosity Tc-rich S-type stars, making it a key constraint on TDU in detailed models. 

\textit{CD $-$30$^{\circ}$4223 (Hen 4-19)}: This star is faint ($\sim2500$ \Lsun), and was identified as Tc-rich in \cite{van_eck_jorissen_tc_1999}. 
It appears to have an LSP of $\sim 200$ or $400$ d which dominates the light curve, with smaller amplitude variability that is challenging to disentangle with the iterative whitening approach due to incorrect fitting of the LSP. A tentative period of 77 d was obtained after some prewhitening steps, though this may still be quite uncertain. However, this period may be valid if we assume the VSX period of 48 d is the 2OM period and the 77 d period to be the 1OM. We estimate an initial mass of $1.61^{+0.23}_{-0.20}$ \Msun, which is consistent within uncertainties across both the 1OM and 2OM model tracks. Of course, this is subject to the uncertainties associated with disentangling the effects of the dominant LSP during light curve analysis; higher cadence light curves of this star may help remedy this to better resolve the shorter pulsation periods. 

Nevertheless, these mass estimates are slightly higher than what one may expect for a star with such a low luminosity, which demonstrates the additional constraint on mass that the pulsation periods impose. A similar effect may be seen for HD 288833 in Figure \ref{fig:g2m_mest_z007}, as the only Tc-rich star in the `RGB and faint AGB' section of the Gaia-2MASS diagram. Its luminosity estimate is very low for a TP-AGB star ($1859 \pm 100$ \Lsun), yet using its period 16.1 d (assigned to the 3OM from its position on the PL diagram) we estimate initial mass of $\sim 1.6 $ \Msun. It is possible that this star is early in the interpulse phase where the luminosity has dipped, and the radial pulsation has reverted to higher order modes. However, this interpretation may be complicated by the possibility that this star is a binary system, with an estimated orbit of the companion is 78 years with an eccentricity of 0.35, though these orbital elements are poorly constrained \citep{jorissen_barium_2019}. HD 288833 was also only recently reclassified from extrinsic \citep{jorissen_tc_1993} to intrinsic in S21 after the detection of Tc. 

\textit{BD $-$18$^{\circ}$2608}: This star was newly classified as Tc-rich in S21, and was estimated as having an initial mass of 1.6 \Msun\ despite its low luminosity at $2400$ \Lsun, for which we find similar values of 2285 \Lsun\ (VOSA) and 2398 \Lsun\ (K10). Unfortunately, its ASAS-SN light curve was affected by saturation, and no reliable period could be found. However, its VSX entry lists a period of $\sim 1700$ d, which is a long period for BD$-$18$^{\circ}$2608's low luminosity. This period would place it beyond the LSP sequence for its $M_{W_{JK_s}}$ ($-7.47$). Upon further investigation of the ASAS and KWS light curves for this star, we find a possible pulsation period of $\sim60$ d after a few prewhitening steps which we attribute to the 1OM. This results in an initial mass estimate of $1.17^{+0.09}_{-0.09}$ \Msun\, using the Z = 0.007 grid which is expected to be comparable to the measured metallicity of [Fe/H] = $-0.31 \pm 0.16$. We also find a cycle of roughly 260 d which puzzlingly lies in between sequences C and D, and the resulting phased light curve appears to have troughs reminiscent of an eclipsing variable. This case may be similar to the stars with primary period between sequences C and D identified in \cite{soszynski_wood_srvs_2013}, though the cause of this variability is (like sequence D) not definitively understood. It is possible that the 287 period is just half of the LSP, in which case it would lie on sequence D. 

These examples illustrate the complications that mass estimates for individual stars in the sample can face, depending on the varying quality and features of their light curves. As a result, the utility of these stars as constraints on the minimum mass for TDU will strongly depend on accurate interpretations of their evolutionary state. The lowest luminosity intrinsic S-type stars tend to be SRVs with complex variability, and while we have obtained reasonable pulsation periods for most, some stars require further monitoring for more accurate measurements. 

\subsubsection{Comparison to initial masses for S stars from Shetye et al. 2021}

\begin{figure}
    \centering
    \includegraphics[width=\linewidth]{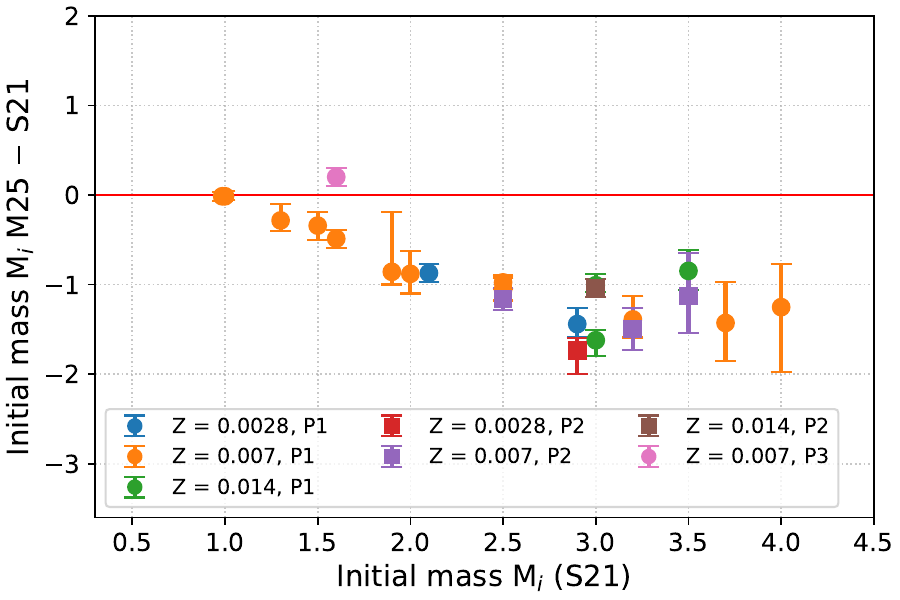}
    \caption{Comparison of initial masses derived in S21 with those derived in the current work. We find that beyond initial masses of 1.5 \Msun, our masses are systematically $\sim 1$ \Msun\ smaller than S21. Symbols indicate the pulsation mode the masses were based on, where circles are the 1OM and squares are the 2OM.}
    \label{fig:compare_s21_masses}
\end{figure}

S21 also make initial and current mass estimates for their intrinsic S-type sample, by comparing their derived stellar parameters on the HR diagram with TP-AGB evolutionary tracks. In Figure \ref{fig:compare_s21_masses}, we compare the initial masses derived with pulsation in this study with the S21 masses. This figure uses the masses derived with the model metallicity closest to that measured in S21 using the same criteria as the metallicity randomisation approach in Section \ref{s:mass_z_alpha_massloss}, and also marks which mode was used to derive the pulsation mass. Stars with M$_i$ $\lesssim$ 1.5 \Msun\ appear to be consistent, but there is a large $1$-$1.5$ \Msun\ offset for stars with S21 masses of $>1.5$ \Msun. The asymmetry for the error bar on CSS 454 ($1.14 ^{+1.06}_{-0.35} M_{\odot}$) is caused by a combination of its large luminosity uncertainty and the lower mass limit of 0.9 \Msun. V915 Aql was measured to have [Fe/H] $= -0.50$, but does not fall within the interpolation grid for the Z = 0.0028 models. Here we adopt the mass obtained from the Z = 0.007 grid, acknowledging the mass may be even lower than the $\sim 1$ \Msun\ we estimate.

We have demonstrated that the luminosities derived in this work and S21 are mostly consistent in Section \ref{fig:compare_s21_vosa_k10_lums}, so the differences likely lie in the method used to derive these masses (HR or PL diagrams), or the stellar models themselves (STAREVOL and Monash). One way that mass estimates may be inflated when using the HR diagram is misattributing a star with the thermal pulse luminosity dip of a higher stellar mass track. Most of the interpulse will be spent near the luminosity maximum, so stars will be more likely to be observed there instead of the lower luminosity phase of the interpulse.

The models used in S21 are sourced from \cite{escorza_hrbarium_2017}, which are computed with the STAREVOL stellar evolution code \citep{siess_starevol_2000}. Figure 9 in \cite{escorza_hrbarium_2017} compares the same STAREVOL HR diagram tracks to results from the Geneva and Padova stellar evolution codes, and demonstrates the differences in evolution tracks between these models. This may be related to differences in model core masses, because the core mass generally governs the luminosity of an AGB star. For the Monash models, the core mass at the first thermal pulse in a 2 \Msun\, Z = 0.014 model is 0.53 \Msun. This is consistent with, though slightly larger than, TP-AGB models in MESA \citep{rees_mesatp_2024} (0.51 \Msun), larger than \cite{weiss_ferguson_agb_2009} (Z = 0.02, 0.478 \Msun), and slightly smaller than \cite{cristallo_agb_2009} (0.548 \Msun).

\begin{figure*}
    \centering
    \includegraphics[width=0.8\linewidth]{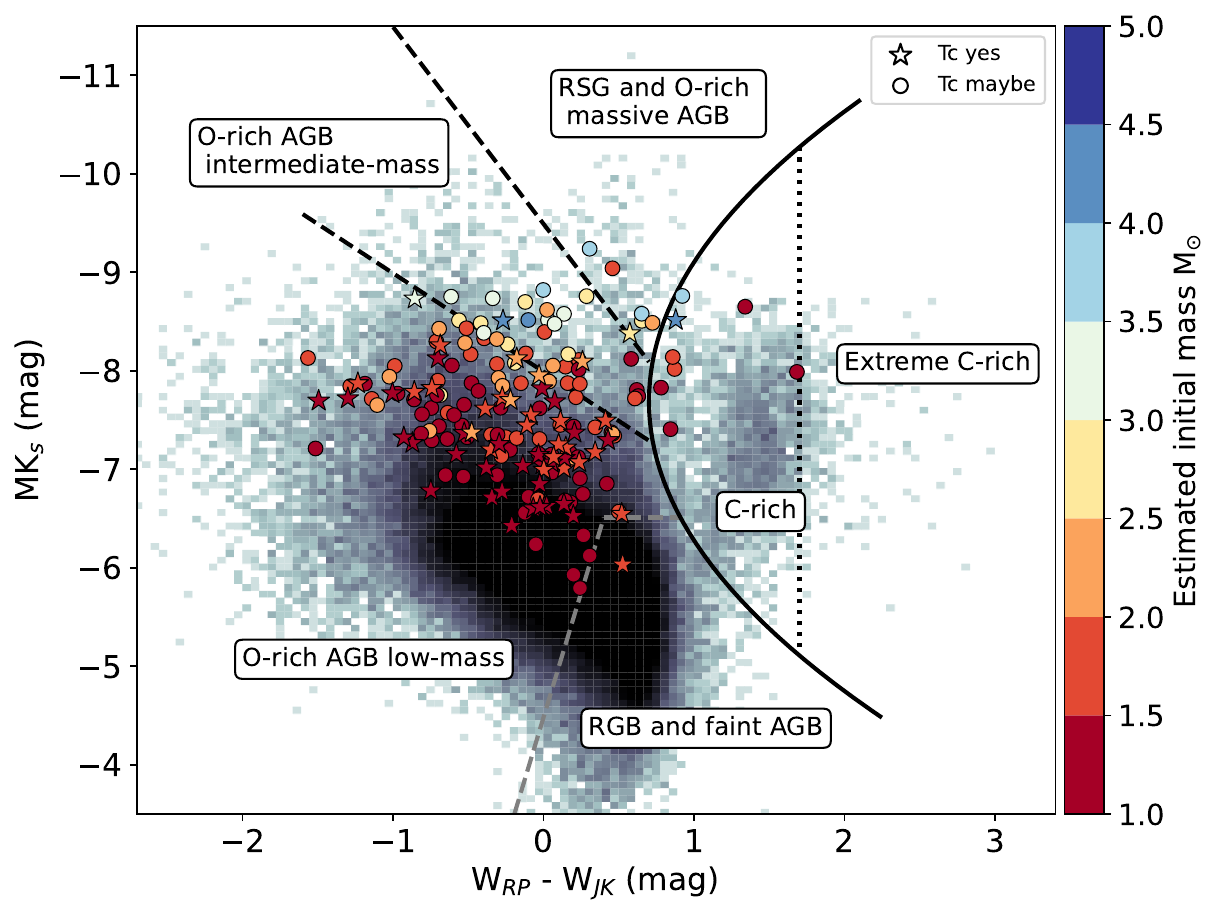}
    \caption{Gaia-2MASS diagram of stars with masses estimated using the first and second overtone mode periods, and models with Z = 0.007. In the background is a 2D density histogram for the ASAS-SN catalogue of LPVs. According to \protect\cite{lebzelter_new_2018}, each section has estimated mass ranges of: O-rich AGB low mass ($\sim$ 0.9 to 1.8 \Msun), O-rich AGB intermediate-mass ($\sim$ 2 to 6 \Msun) and RSG and O-rich massive AGB ($\gtrsim$ 8 \Msun). Note that in the Gaia-2MASS diagram, it is possible we have captured S- or SC-type stars during their transition from the O-rich side to the C-rich side -- this means intermediate mass stars may appear to be in a higher mass category, as they are evolving horizontally from left to right. The Tc-rich star in the RGB and faint AGB section is HD 288833, which has an estimated initial mass of 1.6-1.8 \Msun.}
    \label{fig:g2m_mest_z007}
\end{figure*}

Additionally, S21 verify their masses with the positions of their sample on the Gaia-2MASS diagram. The low-mass O-rich region in which most of the S-type stars lie is defined in \cite{lebzelter_new_2018} to be populated by stars between 0.9 to 1.4 \Msun, sometimes with masses up to 1.8 \Msun\ on the brightest edge. In the S21 Gaia-2MASS diagram, the brightest edge is populated with stars up to $\sim 3.5$ \Msun. This suggests that these masses may be overestimated, at least compared to the population synthesis models from \cite{lebzelter_new_2018}. We construct a Gaia-2MASS diagram with our mass estimates using Z = 0.007 models in Figure \ref{fig:g2m_mest_z007}, separating masses into bins of 0.5 \Msun. Most masses seem to agree with their respective regions, though there still appear to be some stars between 3 and 3.5 \Msun\ that populate the upper edge of the low-mass O-rich region. Some low-mass stars appear to be in the intermediate or massive star regions, but we attribute this to their changing chemistry as identified in Section \ref{s:g2mass_pldiagrams}. Finally, since stars in the range 1 to 3 \Msun\ will all evolve up through the centre of the Gaia-2MASS diagram (0 $<$ WRP$-$WJK $<$ 1) before diverging according to their mass and chemistry, it is difficult to use it as a precise mass diagnostic for less evolved AGB stars. 

\section{DISCUSSION}
\label{s:discussion}

\subsection{Implications of current work}

\subsubsection{Initial masses of intrinsic S-type AGB stars and models of TDU}

A key result of this study is the verification that Tc-rich S-type stars with initial masses below 1.5 \Msun\ seem to exist, which presents a contradiction to the current understanding of TDU in detailed models of TP-AGB stars. The initial mass distribution we derive for the intrinsic S-type stars peaks at roughly $1.4$ \Msun, which is close to the mass inferred from the luminosity function by \cite{abia_characterisation_2022}. This peak mass includes considerations of how metallicity spread may influence the distribution, as well as the uncertainties in mass carried over from the luminosities. The mass distribution we have derived for intrinsic S-type stars should be constrained by the qualitative effects that stellar mass has on TDU. Theoretically, the minimum mass should be limited by the minimum mass of at which TDU occurs and brings up carbon, while the higher mass end of S-type stars may be inhibited by the destruction of carbon due to hot bottom burning (HBB) for stars with $\gtrsim5$ \Msun. Furthermore, the overall distribution will be shaped by the stellar initial mass function, which favours lower mass stars. 

The minimum mass for TDU in current detailed models has previously been calibrated according to observations of the carbon star luminosity function, which appears to indicate that they are made at masses above roughly 1.5 \Msun\ at solar metallicity \citep{groenewegen_carbon_1995,cristallo_agb_2009,pastorelli_smc_2019,pastorelli_lmc_2020}. These models require additional  overshoot at the base of the convective envelope to reproduce carbon star populations, as explored in \cite{kamath_evolution_2012} and references therein. Without any overshoot, models can only produce TDU for initial masses of 2 \Msun\ at solar metallicity \citep[e.g.][]{karakas_agb_carbon_2014}. Recently, \cite{abia_characterisation_2022} estimated that the majority of intrinsic S-type stars are above 1.3 \Msun. However, roughly a quarter of the stars we have estimated a mass for appear to be below 1.3 \Msun, with some of these confirming the low masses of roughly 1 \Msun\ estimated by S21. It is already clear that the minimum mass for TDU must be recalibrated to reproduce S-type stars at these low masses, which will require follow-up from a modelling perspective. The inclusion of the stellar parameters measured in this work (luminosity and pulsation characteristics), as well as published measurements of metallicity and chemical composition will provide additional constraints for the changes to TDU required in detailed models. 

Although it is possible that the low-mass tail of the mass distribution may be alleviated by more accurate mass estimates using metallicity estimates, the intrinsic S-type SRV stars themselves are only a subset of the whole TP-AGB population. The efficacy of these stars as a constraint on TDU would be clarified by an analysis of M-type AGB stars, the likely progenitors of S-type stars. Tc-rich M-type stars are especially interesting cases to probe TDU, as they show signs of Tc despite not yet being enriched enough in carbon to change spectral type \citep{shetye_tcrichm_2022,uttenthaler_interplay_2019}. We have also established that while we have investigated the overtone mode pulsators, pulsation models predict that stars with lower masses spend more time in lower-order pulsation modes. It is important to bear in mind that the intrinsic S-type star population also includes stars with higher initial masses that are just undergoing their earliest TDU events, which are more of a constraint on the lowest luminosity for the earliest TDU event at a given mass, rather than the overall minimum mass for TDU. Models that include overshoot only reach the S-type regime from TDU in late thermal pulses; using the example of the Monash models, a 2 \Msun\ solar metallicity model with overshoot becomes S-type (0.5 $<$ C/O $<$ 1) at the 20th thermal pulse out of 24, becoming C-rich at the final pulse. During most of this time as an S-type star, the pulsation amplitude growth rates predict dominant first overtone mode pulsation. For a similar 1.5 \Msun\ model, the star achieves S-type composition on the 14th pulse of 15, and the majority of these final interpulse phases are expected to be dominant fundamental mode pulsation with some power in the first overtone mode, so the star may be observed as variable type SRa. These examples highlight that it is somewhat puzzling to see low mass ($< 1.4$ \Msun) intrinsic S-type stars as SRVs, when pulsation models predict them to be more likely observed as Miras. Obtaining accurate masses for the Mira variables could reveal intrinsic S-type stars with even lower masses, though precise mass estimates will be complicated by high current mass-loss rates and less certain mass-loss histories.

At the higher mass end roughly of $>5$ \Msun\ (intermediate mass), the production of S-type stars becomes suppressed by the emergence of HBB. Notably, even at these initial masses it is still possible to produce S-type (and even carbon) stars later on the TP-AGB; once mass loss has sufficiently stripped the envelope so that HBB has stopped, dredge-up can continue to enrich the envelope in the final few thermal pulses \citep{frost_brightest_1998}. The intermediate mass S stars in the sample such as NO Aur and V386 Cep may very well be a result of this, and could be interesting subjects to further explore this boundary.

Finally, our results so far have only assumed single star evolution, when in reality observations suggest there is a substantial fraction (40-75\%) of low- to intermediate-mass stars that evolve in binary systems \citep[e.g.][]{moe_distefano_binary_2017}. An important fact to note when using Tc as a marker for the TP-AGB phase is that having Tc does not rule out the star having a lower mass companion. Recent work by \cite{osborn_popsynth_2025} has found that the inclusion of binary evolution in population synthesis models can inhibit the stellar yields of carbon and s-process elements. A lower minimum mass required for TDU may counteract this effect to increase the contribution of TP-AGB stars to galactic chemical evolution, and follow-up calibrations of TDU in population synthesis models using S-type stars may be insightful.

\subsubsection{Composition, pulsation mechanisms and mass-loss of S stars}

The effect of chemical composition on the pulsation properties of LPVs is still an area of active investigation, and our results demonstrate that the measurements of metallicity and C/O ratio are beneficial for more precise mass estimates for these stars.

Models probing the transition from solar-like, stochastically excited oscillations to Mira-like coherent pulsations \citep{cunha_stochastic_2020} suggest that the coherent pulsations likely begin to affect amplitudes from periods between 5-10 d, with the actual transition between the two driving regimes occurring at roughly 60 d. This is interesting in the context of mass-loss, as \cite{mcdonald_zijlstra_60d_2016} identify a sharp onset of dust production (which is linked to mass-loss through dust-driven winds) at a pulsation period of 60 d \citep[also see][]{glass_midir_pmags_2009}. A similar link was noticed by \cite{mosser_plrelations_2013}, who quantitatively demonstrate that the observed maximum variability amplitude, which corresponds to when the surface acceleration due to pulsation is equal to the surface gravity, occurs at a period of 60 d (or $\nu_{\text{max}} \simeq 0.2$ $\mu$Hz). \cite{yu_massloss_2021} also find an additional onset of mass-loss at periods of $\sim100$ d for their collection of LPVs. If these two mechanisms are indeed linked, the pulsation periods we measure for some stars our intrinsic S star sample would lie in a critical stage of activity for pulsation and mass-loss. However, intrinsic S stars generally appear to lie at various pulsation periods and modes, so no definitive links between dredge-up and this onset of mass-loss can be made here.

\subsubsection{Long secondary periods}\label{s:lsps}

Many of these S-type stars have LSPs, which we have decided to classify as distinct from the pulsation periods we measured. This variability may be caused by periodic obscuration by a dusty, close companion \citep{soszynski_binarity_2021} or, as recently proposed in \cite{goldberg_betelbuddy_2024}, modulation of circumstellar material by a close companion. The companion in question is hypothesised to have formerly been a planet that has accreted material from the primary star, and now has roughly the mass of a brown dwarf. However, the origin of this variability is still under debate, with alternative hypotheses such as non-radial oscillation \citep{saio_oscillatory_2015}. The fraction of stars showing LSP behaviour in the present sample ($\sim 38\%$ of the full sample, and $\sim61\%$ of SRVs) is somewhat larger than in prior work, though this may depend on the quality and baselines of the light curves: \cite{wood_macho_1999} find $\sim25\%$ in the LMC with MACHO, \cite{soszynski_oglelpv_2007} estimate 25-50\% of LPVs in the OGLE MC catalogues and \cite{pawlak_lsp_2023} find it to be roughly 27\% of pulsating red giants in the ASAS-SN catalogue of variable stars. Many of the sample stars are pulsating in the first overtone mode, and are around sequence B and C$'$ -- the transition between these sequences has been associated with high-mass loss and the onset of LSPs \citep{mcdonald_trabucchi_plagbwinds_2019}. 

An interesting result for the lowest-luminosity S-type stars is that several are SRVs with LSPs. This appears to be the case in stars such as CD $-30^{\circ}4223$ (Hen 4-19) and CD $-29^{\circ}5912$ (Hen 4-44). Additionally, we find that the candidate bitrinsic S-type star BD $+34^{\circ}1698$ identified in S21 appears to have a LSP in the ASAS-SN light curve. This star has a pulsation period of $86$ d that falls between sequences B and C$'$ of the PL diagram, and a possible LSP of $\sim 700$ d that falls on sequence D. Bitrinsic stars are peculiar stars identified in \cite{shetye_bitrinsic_2020} that show features of both intrinsic (Tc-rich) and extrinsic (large Nb enrichment relative to [Zr/Fe]) stars, and have evidence of white dwarf companions from radial velocity and UV observations \citep{jorissen_barium_2019}. Further radial velocity monitoring of this star, and comparing these measurements to the photometric variability may prove a good test for the companion hypothesis for LSPs. It may also be interesting to explore the mechanism that causes LSPs using the constraints provided by the pulsation characteristics and any derived stellar parameters of the primary stars, though strong evidence of the binary hypothesis is required first. 

\subsection{Qualitative uncertainties}\label{s:qual_uncerts}

\subsubsection{Luminosity uncertainties}

Several studies have demonstrated that Gaia DR3 parallaxes are uncertain for AGB stars and bright stars in general \citep{elbadry_gedr3parallax_2021,andriantsaralaza_distances_2022}. \cite{andriantsaralaza_distances_2022} report that errors of parallaxes are underestimated by more than a factor of 5 for the brightest sources, compared to VLBI measurements. Additionally, the closest AGB stars can have angular sizes larger than the parallax, which can also contribute to further uncertainty. 
The Mira variables in our sample prove to be the most challenging in this aspect, though their stellar parameters may be improved by approaches accounting for the photocentre variability \citep[e.g.][]{beguin_agbparams_2024}.

The luminosity uncertainties for many of our sample stars were actually dominated by the uncertainties in 2MASS photometry, as they are bright stars which often saturate the J and K$_s$ bands. The 2MASS photometry for these saturated stars are estimated from the wings of a saturated stellar image (known as `Read\_3' photometry), which results in a larger uncertainty than for unsaturated sources. Combined with the larger parallax uncertainties for the brighter, more evolved stars, this method also suffers from the larger photometric uncertainty for the closest stars, which may introduce an uncertainty bias favouring close stars with lower intrinsic luminosities. It is possible that this bias may carry through to the mass distributions constructed in Section \ref{s:mass_ests}.

Comparing the PL diagrams of Galactic ASAS-SN and LMC OGLE III samples in Figure \ref{fig:pldiagrams_ogleasassngdr3}, there seems to be an offset to either lower luminosities and/or longer periods that is most obvious for sequence C. Theoretically, stars at higher metallicities may have longer periods, as demonstrated by the model grids, which may be a potential explanation of why Galactic LPVs are shifted to longer periods compared to the lower metallicity LMC. However, this similar shift relative to the PL sequence could be caused by underestimated distances to these stars. Recently, PL relations of SRVs have been used as distance indicators \citep{rau_lpvs_2019,trabucchi_semi-regular_2021,hey_dists_2023}, which may present an alternative avenue for obtaining accurate distances to AGB stars pulsating in the overtone modes. However, the impacts of metallicity on the PL relations still requires further investigation. It may prove informative to compare the Gaia DR3 (and upcoming DR4) parallax distances to these stars with a method similar to that of \cite{hey_dists_2023}. 

\subsubsection{Light curve analysis uncertainties}

There are various caveats to consider in the light curve analysis process, especially for the semi-regular variable stars. Interpretation of the Lomb-Scargle periodogram can be complicated by the fact that the highest peak may not indeed be the true peak, and instead due to an effect of aliasing. False peaks can be introduced from the convolution of the true peak with the observing window, especially for scattered light curves such as those in this work. Since ASAS-SN uses multiple cameras at several locations across the globe, the effect of seasonal aliasing appears to usually be minimal -- the strongest alias periods often appear at $\sim 1$ yr. It is also perhaps difficult to disentangle an $n$ harmonic peak from aliasing and the actual first-overtone mode peak for fundamental mode pulsators, as they both would appear as half of the fundamental mode period. Harmonic peaks of the true frequency can appear when the periodic signals are not strictly sinusoidal, which is often the case for high amplitude fundamental mode pulsators (Miras). The above caveats may have lead to some contamination in our period results. 

The iterative whitening approach in our study makes the assumption that the frequencies removed from the light curve are sinusoidal when fitting. This model is not applicable for stars with LSPs, as these variations are not strictly sinusoidal and result in poor fits. An alternative approach to these light curves may be Gaussian process (GP) regression \citep[eg.][]{foreman-mackey_celerite_2017,scicluna_pgmuvi_2023}. This approach would be advantageous for SRVs: a well motivated model could more flexibly fit variability from both pulsations and LSPs, will not suffer from window aliasing issues like the Lomb-Scargle periodogram, provide probabilistic results for periods and amplitudes and potentially fit for multiple photometric bands simultaneously. 

\subsubsection{Pulsation mode uncertainties}

Precise mode assignment for the pulsation periods of SRVs also requires further investigation. There are finer structures within the PL diagram for long-period variables, as demonstrated by \cite{wood_pulsation_2015} with the OGLE III catalogue for the Magellanic Clouds. It may be a limitation to only consider the radial modes for the SRVs, since it has been demonstrated that they show non-radial pulsations. For example, the assumption that the measured first overtone mode periods are of the radial mode may not be correct, and the dominant mode may be in fact the quadrupole mode ($l=2$) as found in \cite{yu_lpv_2020}. Similarly, this paper finds that the second overtone modes may be dominated by the dipole ($l=1$) mode. The quadrupole and dipole mode periods are slightly longer than the radial ($l=0$) mode periods, which may induce a slight offset relative to the models, leading to underestimated stellar masses. This offset is noted in \cite{wood_pulsation_2015}, who increased their model periods by an empirically derived $\Delta \log P \sim 0.043$ when estimating stellar masses, though they conclude that the first overtone mode is instead dominated by the $l=1$ mode, which is longer than the $l=2$ mode. Nevertheless, we test this same shift for the first overtone mode periods and find that our masses would generally shift upwards by 0.1 \Msun, closer to the theoretical lowest mass for TDU of $\sim1.5$ \Msun. 

On the other hand, \cite{mosser_plrelations_2013} found that the radial modes dominate for longer pulsation periods. Recent results from 3D pulsation models of AGB stars also suggest that non-radial modes may not be so important for most of the sample stars. \cite{ahmad_multimode_2025} find that for higher model luminosities (log $L/L_{\odot}> 3.5$) the non-radial modes appear to blend with the radial modes. For lower luminosities they find that the first overtone mode is dominated by the dipole mode, and the second overtone by the quadrupole mode, which is an inversion of the results from \cite{yu_lpv_2020}. In the current context, only 20\% of the SRVs have log $L/L_{\odot} < 3.5$, so the radial mode may be dominant for the rest. The findings from \citeauthor{ahmad_multimode_2025} also suggest that the appearance of non-radial modes are more an effect of higher mean densities, which raises the question of how non-radial and radial modes vary across the thermal pulse cycle. Perhaps non-radial modes in TP-AGB stars will be more likely observed during the low-luminosity phase of the interpulse, when the star has contracted following a thermal pulse. There are new large grids of stellar oscillations on the AGB computed with MESA and GYRE \citep{joyce_rhya_2024} which may allow for this kind of analysis, as GYRE can also calculate non-radial modes. Further investigation of non-radial pulsation on the TP-AGB would be beneficial for precise mode assignment and mass estimates.

\section{SUMMARY}
\label{s:conclusion}

In this study, we have made estimates for the initial masses of a large sample of intrinsic S-type AGB stars using their pulsation characteristics. We derive masses using the overtone mode periods of the semiregular variables in the sample, in combination with new luminosities derived from Gaia DR3 parallax distances and results of published linear pulsation models and a tailored grid of detailed stellar evolution models. These masses are used to construct initial mass distributions for Galactic S-type AGB stars. We find that this mass distribution peaks at approximately $1.4$ \Msun\ depending on the metallicity spread, and includes stars with initial masses even down to $\sim1$ \Msun. These stars clearly appear to be below the lowest mass expected for the onset of TDU in detailed stellar models of $1.5$\Msun. Another key component in constraining the minimum mass required for TDU is the mass distribution of M-type TP-AGB stars that are yet to, or may not, experience TDU -- we aim to derive this in future work.

Our final results were based on the overtone pulsation modes, as we found that the observed stellar parameters matched detailed models well for these stars. However, the more evolved fundamental mode pulsators proved difficult to obtain reliable mass estimates for. This is due to observational limitations related to uncertain parallax distances and saturated photometry, as well as model uncertainties involving fundamental mode pulsation and mass-loss. Further studies on using the pulsation characteristics of semiregular variable AGB stars to constrain their masses may be promising, since we can exploit multiple radial modes of pulsation. We emphasise that initial mass estimates are also model-dependent, as the treatment of core helium burning, convective overshoot, and mass-loss prescriptions require careful consideration when modelling these stars. 

The complexities of pulsation on the TP-AGB requires further investigation, given the non-linear behaviour of pulsation at larger amplitudes and the uncertain dependence of pulsation on chemistry. Additionally, exploration of the measurement and modelling of non-radial oscillations in AGB stars could provide a stronger foundation for more accurate and precise mode assignment and stellar parameter estimation for these stars. Combining the abundance of long baseline light curves currently available with large upcoming/future datasets (eg. OGLE IV Galactic semiregular variables, \citealp{iwanek_ogle_2022}, the Legacy Survey of Space and Time (LSST) at the Vera C. Rubin Observatory, \citealp{hambleton_rubin_2023}) will likely provide more insights into the pulsation characteristics of TP-AGB stars, and a better understanding of their complex evolution.

\begin{acknowledgement}
    
The authors thank the anonymous reviewer for their constructive comments, Taïssa Danilovich for early comments, and advice from Daniel Hey and Daniel Huber. 

YLM receives support from the Australian Government Research Training Program. SWC acknowledges federal funding from the Australian Research Council through a Future Fellowship (FT160100046) and Discovery Projects (DP190102431 and DP210101299). Parts of this research were funded by the Australian Council Centre of Excellence for All Sky Astrophysics in 3 Dimensions (ASTRO 3D), through project number CE170100013.

This work has made use of data from the ASAS-SN survey, the ASAS-SN Catalogue of Variable Stars, the OGLE III Catalogue of Variable Stars, and data products from the Two Micron All Sky Survey (2MASS; M. F. Skrutskie et al. 2003), which is a joint project of the University of Massachusetts and the Infrared Processing and Analysis Center/California Institute of Technology, funded by the National Aeronautics and Space Administration and the National Science Foundation. This work has made use of data from the European Space Agency (ESA) mission Gaia (https://www.cosmos.esa.int/gaia), processed by the Gaia Data Processing and Analysis Consortium (DPAC, https://www.cosmos.esa.int/web/gaia/dpac/consortium). Funding for the DPAC has been provided by national institutions, in particular the institutions participating in the Gaia Multilateral Agreement.

\begin{sloppypar}
    This research has used data, tools or materials developed as part of the EXPLORE project that has received funding from the European Union’s Horizon 2020 research and innovation programme under grant agreement No 101004214. This publication makes use of VOSA, developed under the Spanish Virtual Observatory (https://svo.cab.inta-csic.es) project funded by MCIN/AEI/10.13039/501100011033/ through grant PID2020-112949GB-I00. VOSA has been partially updated by using funding from the European Union's Horizon 2020 Research and Innovation Programme, under Grant Agreement nº 776403 (EXOPLANETS-A). This research has made use of the VizieR catalogue access tool, CDS, Strasbourg, France (DOI : 10.26093/cds/vizier). The original description of the VizieR service was published in 2000, A\&AS 143, 23. 

    \textit{Code}: \texttt{Astropy} \citep{astropy:2013,astropy:2018,astropy:2022}, \texttt{balmung} \citep{hey_dscutikepler_2021}, \texttt{giraffe} \citep{trabucchi_modelling_2019}, \texttt{matplotlib} \citep{Hunter:2007}, \texttt{NumPy} \citep{harris2020array}, \texttt{pandas} \citep{reback2020pandas}, \texttt{pyasassn} \citep{hart_skypatrol2_2023}, \texttt{SciPy} \citep{2020SciPy-NMeth}.
\end{sloppypar}

\end{acknowledgement}

\section*{Conflicts of Interest}

None.

\section*{Data Availability Statement}

Tables of the initial sample, results for measured periods, luminosities and mass estimates are made available via PASA Datastore. Additional AGB models computed for this paper are available upon request to the authors.


\bibliography{library}

\appendix

\newpage

\section{Additional luminosity diagrams}

\begin{figure}
    \centering
    \includegraphics[width=\linewidth]{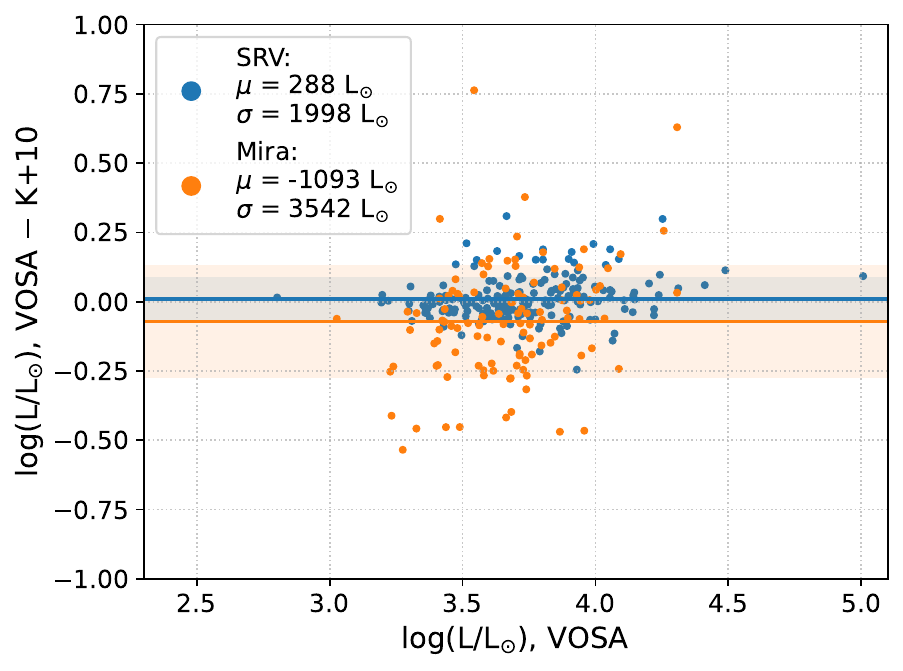}
    \caption{Comparison of luminosities estimated with the \protect\cite{kerschbaum_bolometric_2010} bolometric corrections and VOSA SED fits for the semiregular and Mira variables. The mean and standard deviation of the residuals for each group reveals larger scatter and offset in the Mira variables ($\sigma \sim 3500$ \Lsun), while the semiregular variables are more consistent. VOSA luminosities for Miras are dimmer than those calculated with the K10 BC, which may be caused by the larger photometric variability of these stars in the visual bands.}
    \label{fig:lum_residuals_vartype}
\end{figure}

\begin{figure}
    \centering
    \includegraphics[width=\linewidth]{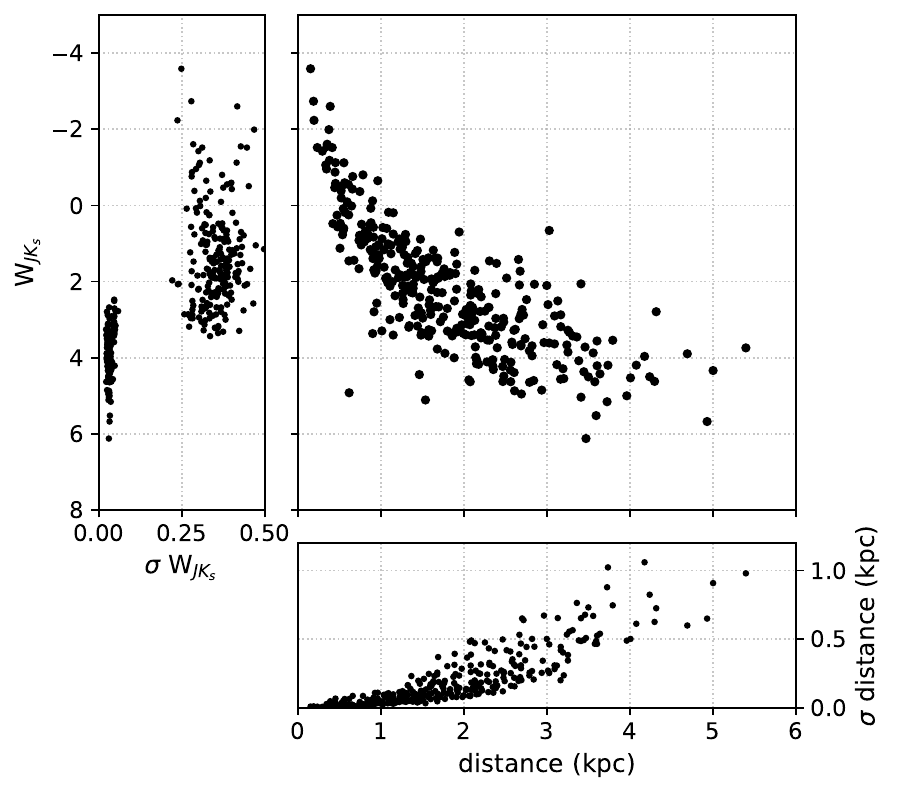}
    \caption{$W_{JK_s}$ index against the geometric parallax distance for the intrinsic S star sample, with associated uncertainties. The photometric uncertainty has a two distinct groups, which is a result of the different treatment of saturated photometry for the brightest stars. The distance uncertainties increase significantly beyond roughly 1 kpc, up to $\sim$ 25\%.}
    \label{fig:enter-label}
\end{figure}

\section{Example VOSA SED Fit}

Example fit to SED using VOSA.

\begin{figure}
    \centering
    \includegraphics[width=\linewidth]{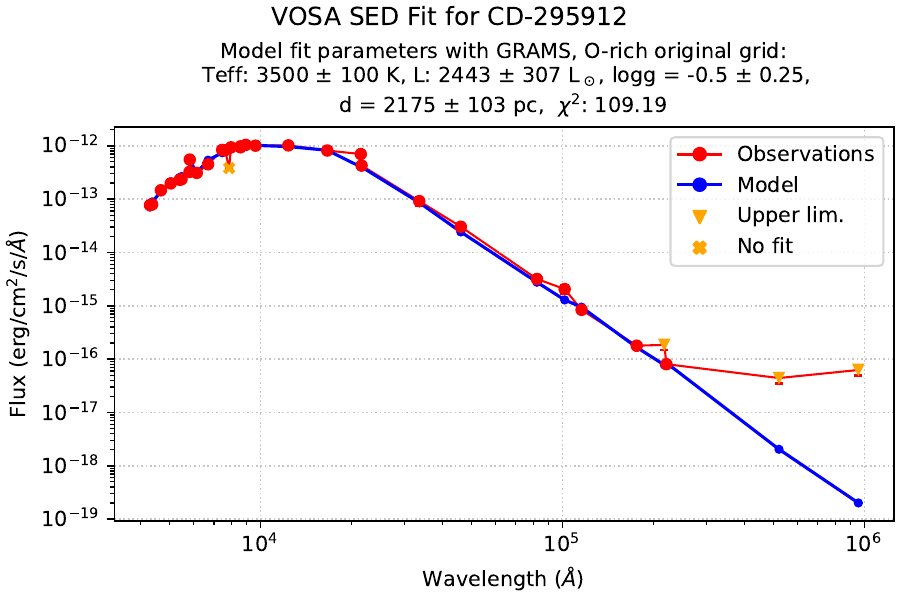}
    \caption{Example VOSA SED fit for CD$-$29$^{\circ}$5912 (Hen 4-44).}
    \label{fig:egvosafit}
\end{figure}

\section{Additional light curves}\label{s:appendix_v915aql}

Light curves for V915 Aql, showing the main variability period of 80.3 days.

\begin{figure}
    \centering
    \includegraphics[width=\linewidth]{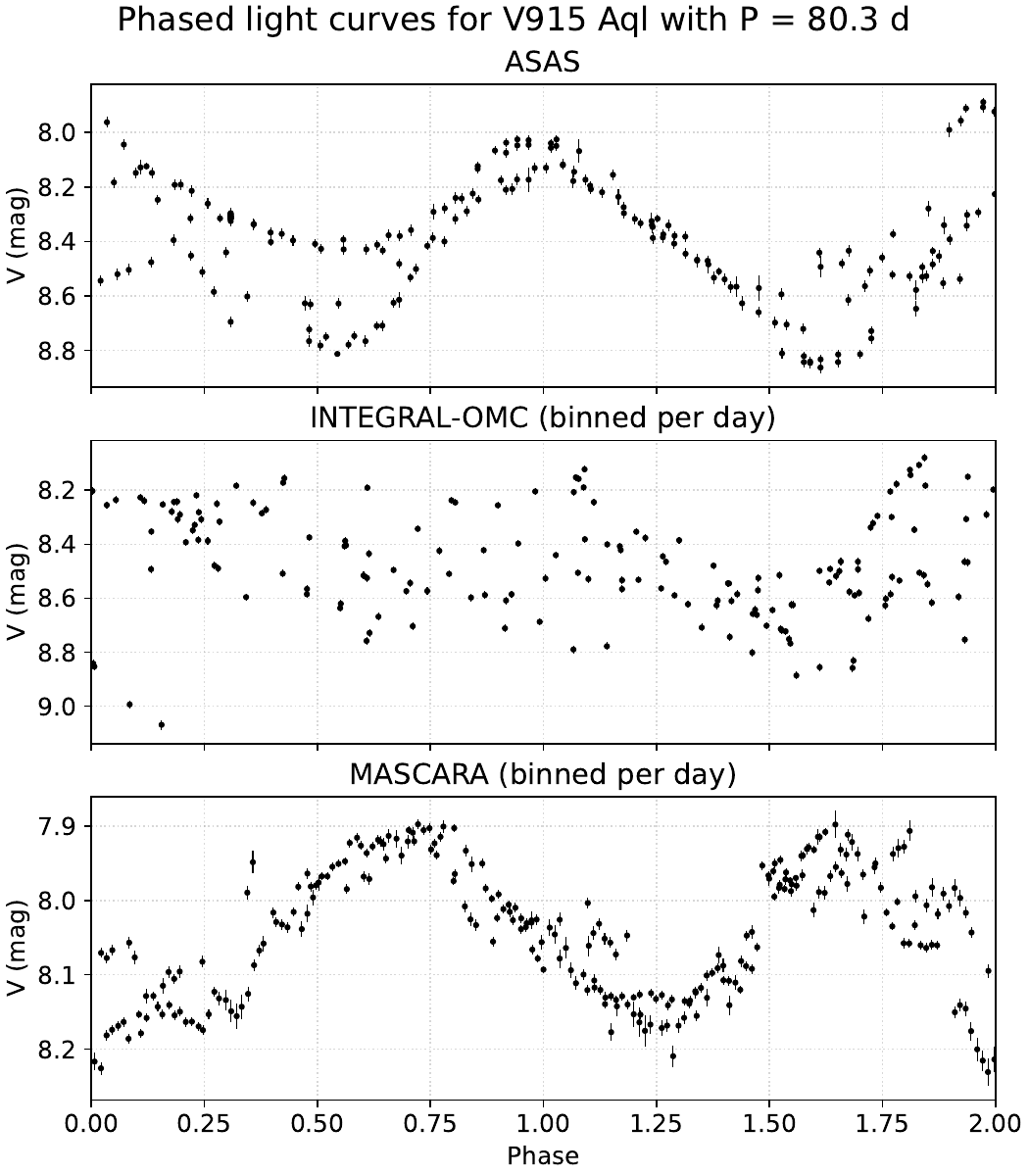}
    \caption{Phased light curves for V915 Aql, using data from ASAS, INTEGRAL-OMC and MASCARA showing two cycles. The INTEGRAL-OMC and MASCARA data are binned per day, and we use the VSX period of 80.3d to fold the light curves. This period is consistent with the ASAS and MASCARA light curves, but less so for the INTEGRAL-OMC data.}
    \label{fig:v915aql_phased}
\end{figure}

\section{Mass distributions for pulsation modes and metallicities}

\begin{figure*}
    \centering
    \includegraphics[width=\linewidth]{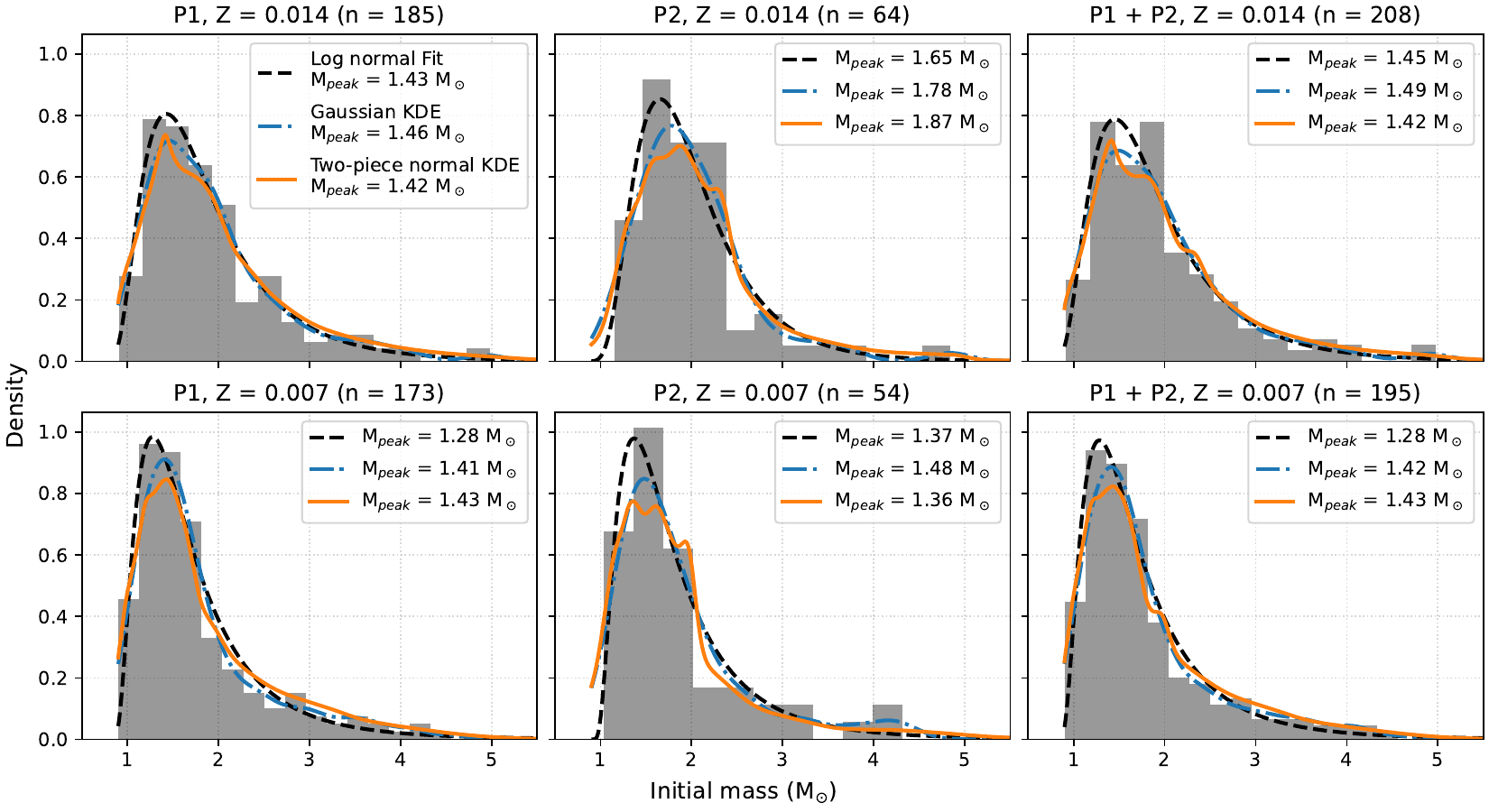}
    \caption{Histograms and distributions of mass estimates for first (P1) and second (P2) overtone mode pulsators, using models with Z = 0.014 and 0.007.
    We also note the peak mass of each distribution according to a log normal fit, a Gaussian kernel density estimation and a two-piece normal KDE. The Gaussian KDE uses the Silverman formula to estimate the bandwidth, while the two-piece normal KDE uses the upper and lower mass uncertainties as $\sigma$ for each side of the kernel.
    A lower assumed metallicity shifts the mass distribution to lower masses by up to $\sim 0.2$ \Msun. The second overtone mode pulsators were found to have larger masses on average, which is perhaps due to higher mass stars spending more time in higher overtone modes over the TP-AGB.}
    \label{fig:mass_ests_all}
\end{figure*}

The KDE mass distributions for each combination of pulsation mode and model metallicity.

\begin{figure}
    \centering
    \includegraphics[width=\linewidth]{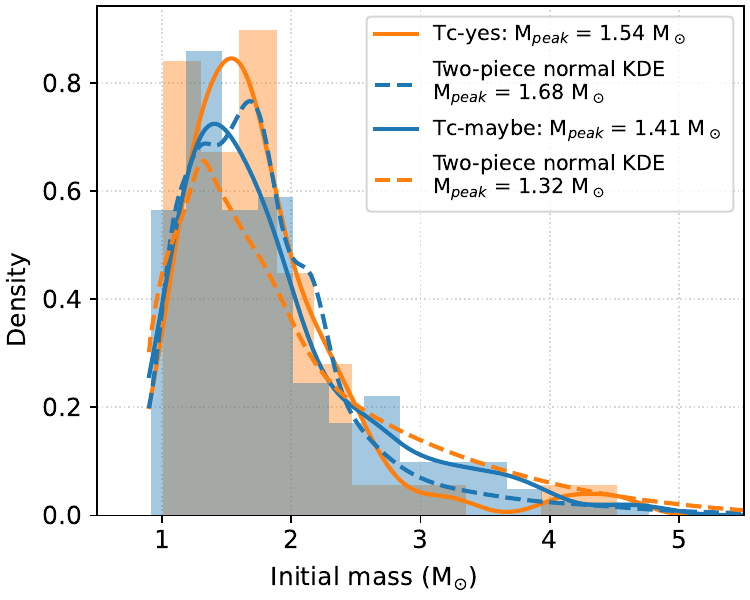}
    \caption{Histograms and Gaussian KDEs for the mean mass estimates, but split between Tc-yes and Tc-maybe subsamples.}
    \label{fig:massest_tcytcm}
\end{figure}

\section{Fundamental mode models and observations}

\begin{figure}
    \centering
    \includegraphics[width=\linewidth]{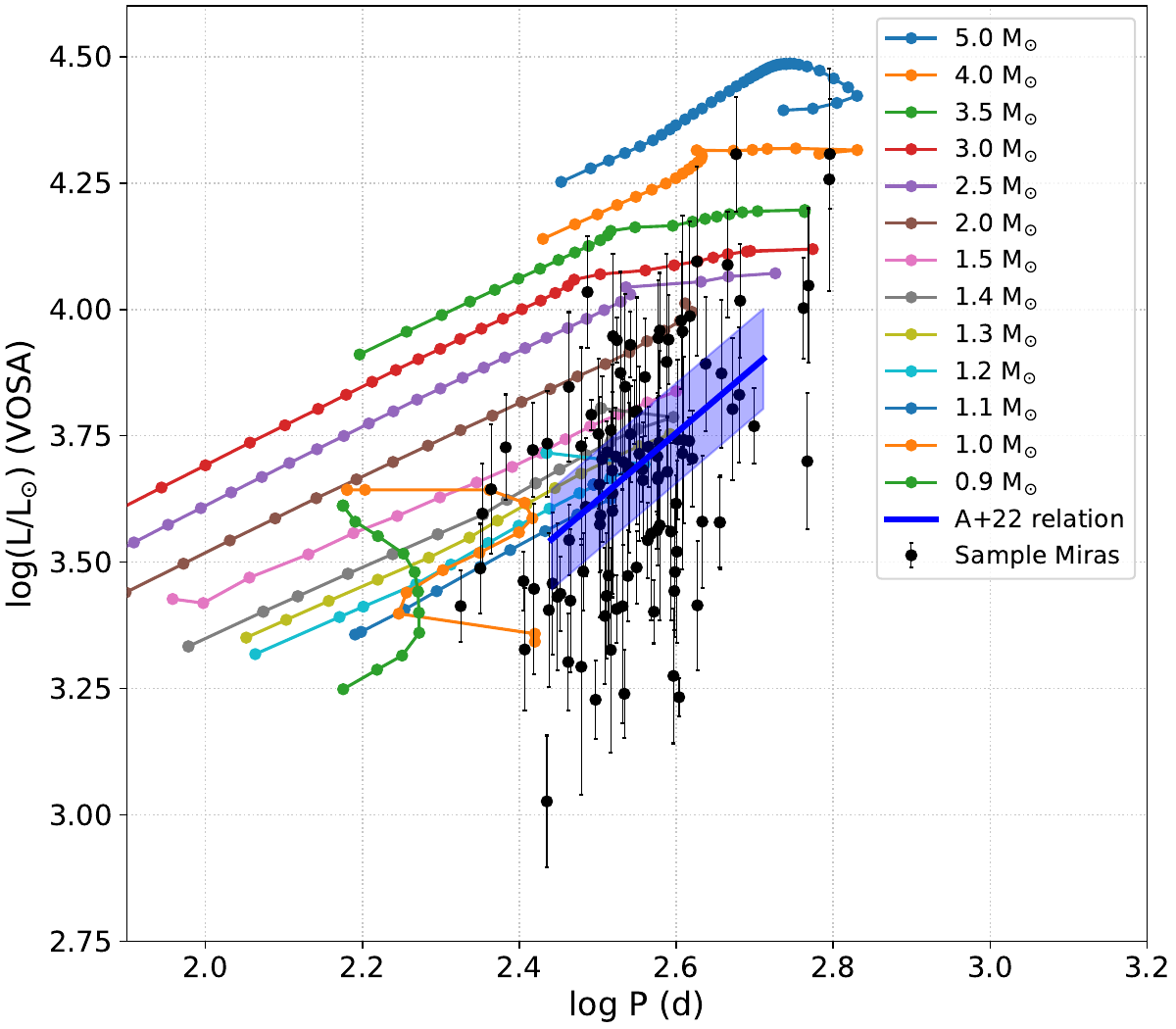}
    \caption{Theoretical period-luminosity diagram for the non-linear fundamental mode pulsation period from \protect\cite{trabucchi_modelling_2021}, with sample Mira variables and the \protect\cite{andriantsaralaza_distances_2022} O-rich Mira period-luminosity relation. Many of the sample stars lie at luminosities below the model tracks, with significant scatter. The \protect\cite{andriantsaralaza_distances_2022} relation also appears to lie at lower luminosities (or longer periods) compared to the model tracks.}
    \label{fig:fm_model_a22_sample}
\end{figure}

\end{document}